\documentclass[twocolumn,pre,superscriptaddress]{revtex4-1}

\usepackage{bbm}
\usepackage{graphicx}
\usepackage{amsmath,amsfonts,amssymb}
\usepackage{multirow}
\usepackage{rotating}
\usepackage{pseudocode}

\newcommand{\SI}{\textit{SI Appendix}}

\begin{document}
\title{Environmental structure and competitive scoring advantages in team competitions}

\author{Sears Merritt}
\email[]{sears.merritt@colorado.edu}
\affiliation{Department of Computer Science, University of Colorado, Boulder, CO 80309}

\author{Aaron Clauset}
\email[]{aaron.clauset@colorado.edu}
\affiliation{Department of Computer Science, University of Colorado, Boulder, CO 80309}
\affiliation{BioFrontiers Institute, University of Colorado, Boulder, CO 80303}
\affiliation{Santa Fe Institute, 1399 Hyde Park Rd., Santa Fe, NM 87501}

\begin{abstract}
In most professional 
sports, the structure of the 
environment is 
kept neutral so that 
scoring imbalances may be attributed to differences in team skill. It thus remains unknown what impact structural heterogeneities can have on scoring dynamics and producing competitive advantages. 
 Applying a 
generative model of scoring dynamics to roughly 10 million 
team competitions drawn from an 
online game, we quantify the relationship between a competition's structure and its 
scoring dynamics. Despite wide structural
variations, 
we find the same three-phase pattern in the tempo of events observed in many 
sports. 
Tempo and balance are highly predictable from a competition's structural features alone 
and teams exploit environmental 
heterogeneities for sustained competitive advantage. 
The most balanced competitions are associated with specific environmental heterogeneities, not from equally skilled teams
. These results shed new light on the principles of balanced competition, and illustrate the 
potential of online game data for investigating social dynamics and competition.
\end{abstract}

\maketitle

Professional team sports are a rich and relatively controlled domain through which to investigate fundamental questions in both the dynamics within and across competitions between groups, and the factors that determine competitive outcomes~\cite{reed:hughes:2006,bourbousson:seve:mcgarry:2012}. With many possible actions and many possible payoffs, such games are a kind of dynamical competition~\cite{galla:farmer:2013}, in contrast to the strategic interactions of classic game theory~\cite{myerson:1997}.
A distinguishing feature of most such competitions is their structurally homogeneous or ``level'' playing field, which allows differences in team scores to be attributed to one team being relatively more skilled than another, or, if the difference is small, to chance events~\cite{denrell:2004,herbrich:minka:graepel:2007}.

It thus remains unknown what impact structural heterogeneities, like an irregular playing field, variations in rules, or differences in resources, may have on a competition's internal dynamics. Heterogeneities may produce structural competitive advantages~\cite{barney:1991}, allowing a team to perform above its skill level by exploiting these environmental irregularities. In fact, the roles in shaping competition dynamics and outcomes of skill, structure, and chance remain highly controversial, both in sports~\cite{bennaim:etal:2012} and in other types of social competition~\cite{barney:1991,salganik:etal:2006,denrell:liu:2012}. A better understanding of these principles would inform the design of novel competitive environments~\cite{michael:chen:2005,nisan:etal:2007}, and could shed light on competition dynamics in other domains, such as ecology and evolutionary biology~\cite{boucher:1985}, political conflict~\cite{myerson:1997} and economics~\cite{bowles:2006}.

Online games present a novel approach to investigate these questions. Such games encompass a broad and growing variety of relatively controlled competitions, played by hundreds of millions of individuals~\cite{esa:2011} and producing large quantities of detailed observational data. We study a unique data set drawn from the popular online game \textit{Halo} (see \SI), a kind of virtual team combat, which contains nearly 1 billion scoring events across roughly 10 million diversely structured team competitions. Each of these competitions is roughly independent, such that team memberships are substantially randomized and no acquired resources are carried to the next competition. This property thus mitigates the confounding effects of cross-competition correlations present in professional sports and allows us to study how structural variations shape competition dynamics and outcomes.

We partition these competitions according to their particular environmental structure, competition rules, resource quality and difference in team skill, and characterize their scoring dynamics via a probabilistic model. The resulting model parameters provide a compact representation of the associated competitive dynamics, and serve as targets to be explained by variation in a competition's structural features.

Despite wide variation, structure has a modest impact on the tempo of events, but a large impact on the scoring balance, i.e., the difference in team scores. Additionally, the rate of scoring events over time exhibits the same three-phase pattern observed in professional sports~\cite{gabel:redner:2012}. Overall, structural features alone are highly predictive of overall competition tempo, the range of competitive scoring advantages, and ultimate predictability of the competition's outcome. Like business firms competing in the marketplace~\cite{barney:1991}, teams generally exploit environmental and resource heterogeneities for sustained competitive advantage. However, contrary to the pattern of professional sports, the most balanced competitions---those with narrow margins of victory---arise from specific environmental heterogeneities, not from equally skilled teams competing in homogeneous environments. These results illustrate the rich potential of online game data for investigating social dynamics and competition~\cite{szell:lambiotte:thurner:2010}, clarify the role of chance when teams are well matched, and point to specific design principles for balanced competitions.

\section*{Results}

\noindent \textbf{Quantifying competition dynamics}. 
We first introduce the notion of an ``ideal'' competition, in which perfectly matched teams play on a level field with no exploitable features. Such a competition's outcome is thus determined solely by the occurrence and accumulation of chance events, e.g., accidents, miscalculations, and events outside direct control. In this way, the highly strategic and carefully motivated actions of equally skilled teams will effectively produce purely stochastic dynamics.

These dynamics can be described by a particularly simple stochastic process~\cite{brzezniak:zastawniak:2000}. Scoring events occur infrequently and independently, and their pattern follows a Poisson process with rate $\lambda_{0}$---a common assumption in quantitative analysis and modeling of professional sports~\cite{gabel:redner:2012,thomas2007inter,heuer2012does}. Given a scoring event occurs, a fair coin determines which team accrues points from it. The difference in scores between teams thus follows an unbiased random walk, and scoring overall follows an equiprobable or balanced Bernoulli scheme.

Real competitions, with heterogeneous structure or skill differences, will deviate from this ideal. We capture these deviations through a generalized model, which may be fitted directly to scoring data and whose parameters quantify the size and character of the non-ideal patterns. We then investigate the extent to which the observed non-ideal patterns can be predicted from variation in competition structure.

We assume a competition between teams $r$ and $b$, and we let $s_{r}(t)$ denote team $r$'s cumulative score at an intermediate time $t<T$. The probability that $r$'s score increases at time $t$ is given by the joint probability of a scoring event occurring at $t$ and of $r$ scoring it. Letting these probabilities be independent yields
\begin{align}
\Pr(\Delta s_{r}(t) > 0) = \Pr(\Delta s_{r}>0\,|\,\theta,\textrm{event})\Pr(\textrm{event at }t\,|\,\theta\,) \nonumber \enspace ,
\end{align}
where $\theta$ parameterizes the non-ideal patterns.

Scoring events occur infrequently and independently, and are now produced by a simple non-stationary point process, in which the arrival of events varies linearly with time:
\begin{align}
\Pr(\textrm{event at }t\,|\,\lambda_{0},\alpha\,) & =  \lambda_{0}+\alpha\,t \enspace . \nonumber
\end{align}
The base or background rate is given by $\lambda_{0}$ and $\alpha$ parameterizes the non-stationarity, e.g., increasing ($\alpha>0$) or decreasing ($\alpha<0$) tempo. When $\alpha=0$, we recover the ideal case of a Poisson process with rate $\lambda_{0}$.

The score of a team follows a general Bernoulli process. Given a scoring event, points are awarded to team $r$ with some probability that is fixed for this competition, but which may vary between competitions
\begin{align}
\Pr(\Delta s_{r}>0\,|\,\textrm{event}) & =  c \enspace , \nonumber
\end{align}
and otherwise, they are awarded to team $b$. This scoring bias $c$ is a probabilistic measure of $r$'s competitive advantage over $b$, e.g., from a difference in skill or from exploitable features of the competition. When $c=1/2$, we recover the ideal case of a balanced Bernoulli process, while deviations produce the more lopsided trajectories associated with non-ideal dynamics. 

Across competitions with the same structure, different pairs of teams will exhibit different competitive advantages. Thus, the natural explanatory target is the distribution of the scoring imbalances $\Pr(c)$, whose natural form is a symmetric Beta distribution~\cite{bishop:2006} (see \SI), the conjugate prior for the Bernoulli process. The result is a one-parameter model that quantifies the overall variability in competitive advantages across a set of competitions. The ideal case of perfectly matched teams and scoring differences due only to chance events occurs at $c=1/2$, which is recovered in the limit of $\beta\to\infty$. Smaller values of $\beta$ indicate less balanced and thus more predictable scoring dynamics across the set.

We supplement this parametric approach with a non-parametric measure of non-ideal behavior: the predicability of the winner from a partially unfolded competition. Having observed the first $k$ scoring events, predicting the winning team is a kind of classification task, which we formalize as a Markov chain on the sequence of team scores (see \SI). For two-team competitions, the probability that team $r$ wins, given current scores $s_r$ and $s_b$, is
\begin{align}
\Pr(r \textrm{ wins} \,|\, s_r, s_b) = \,& \Pr(r \textrm{ wins} \,|\, s_r+1, s_b)\cdot \hat{c} \,\, + \nonumber \\
 &  \Pr(r \textrm{ wins} \,|\, s_r, s_b+1)\cdot (1-\hat{c}) \enspace , \nonumber
\end{align}
where ${\hat c} = s_r/ (s_r+s_b)$ estimates $r$'s competitive advantage. After each event, the classifier predicts as the winner the team with the greatest estimated odds-to-win, and its accuracy is measured by the AUC statistic~\cite{bradley:1997}, the probability of choosing the correct winning team.

The AUC versus $k$ provides complete information about a competition's predictability but is not amenable to our subsequent analysis. We instead use a point measure $\rho$, defined as the ratio of the Markov classifier's AUC to that of an ideal competition ($c=1/2$), when 20\% of the competition has unfolded. A value of $\rho>1$ indicates that the competition outcomes are more predictable than in the ideal case.

\begin{figure*}[t!]
\begin{center}
\vspace{-7mm}
\includegraphics[scale=0.40]{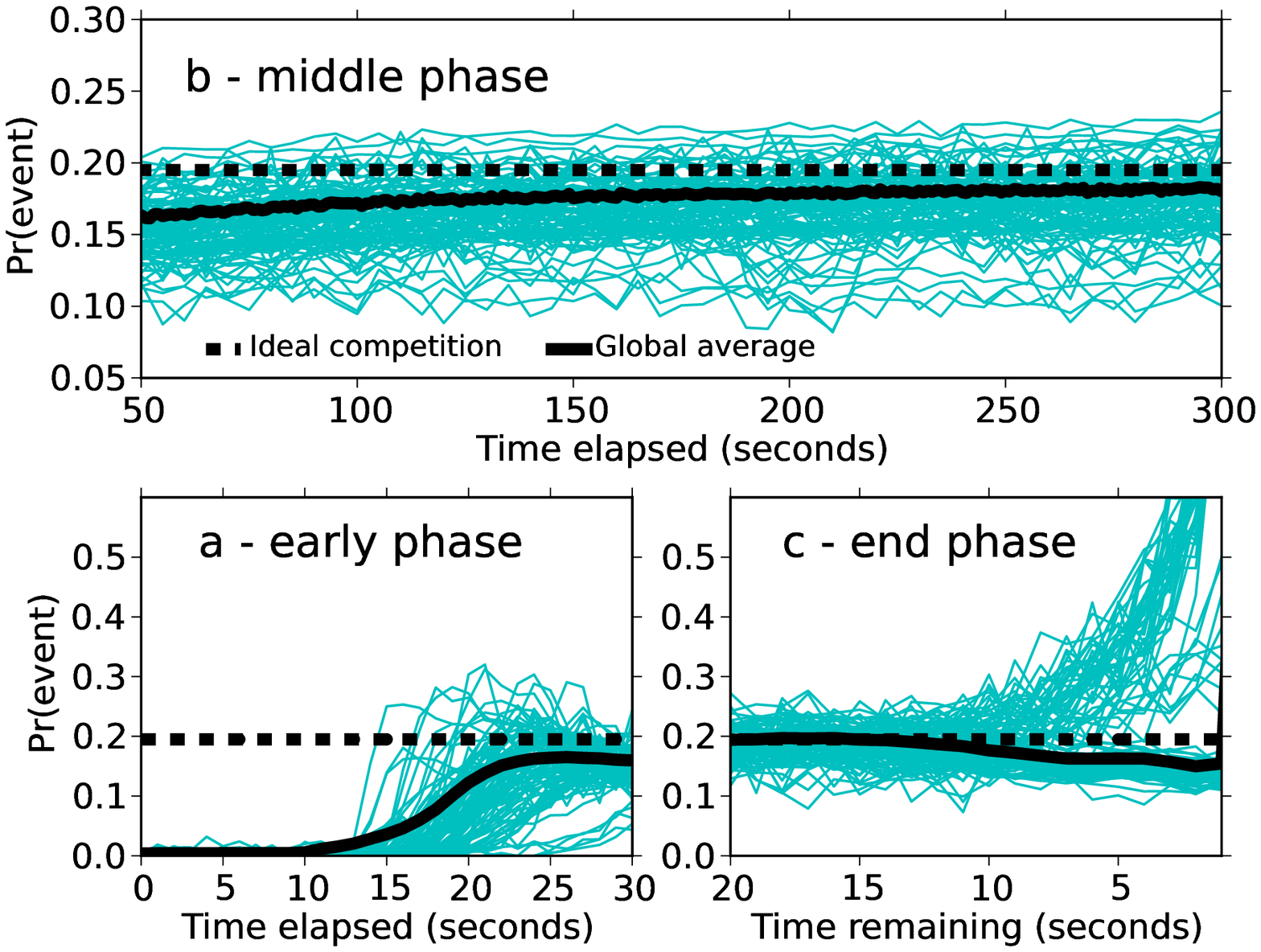}
\includegraphics[scale=0.40]{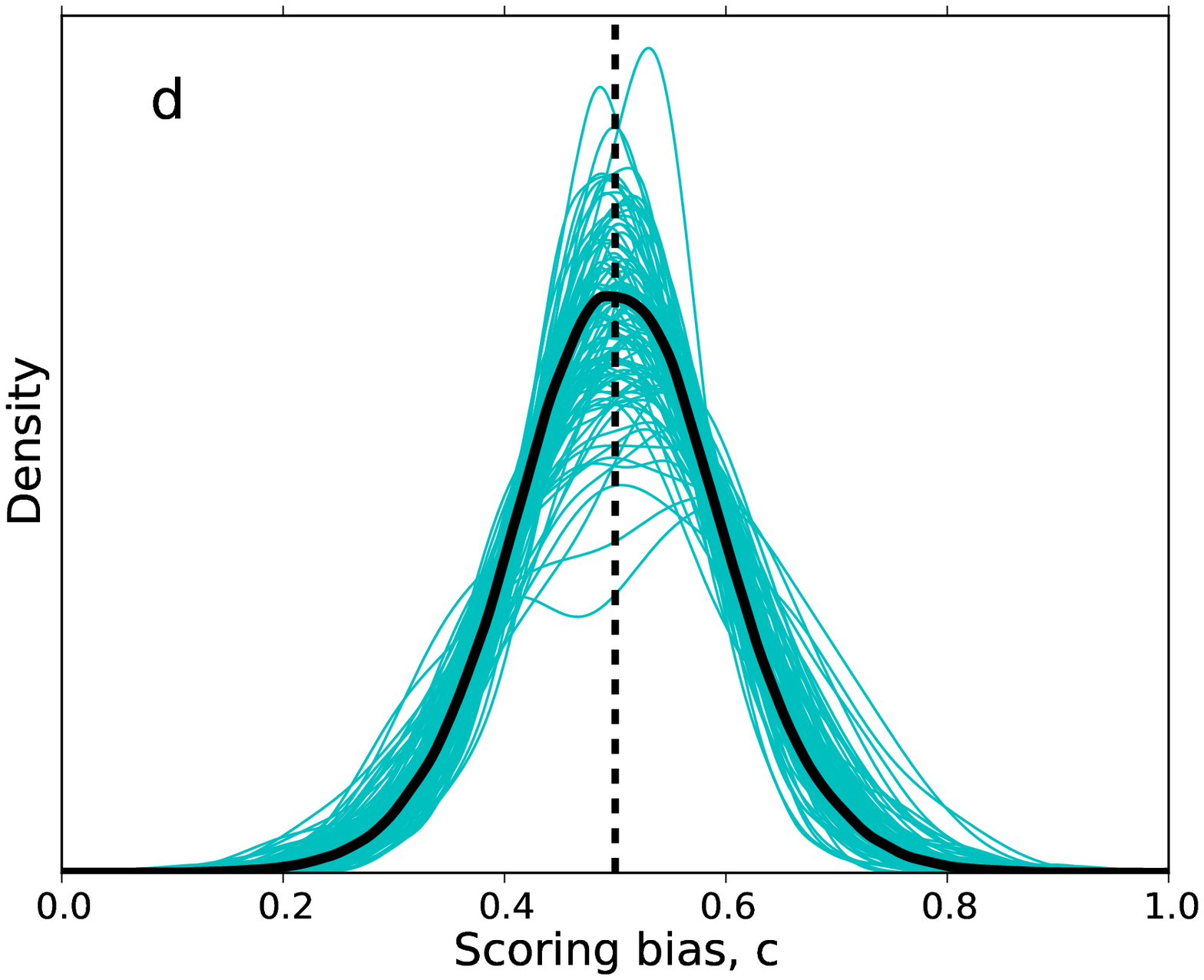}
\caption{Patterns in tempo and score dynamics. For each of 125 competition types, the probability of a scoring event at time $t$, in the (\textit{A}) early, (\textit{B}) middle and (\textit{C}) end phases of a competition; and (\textit{D}), the distributions of the probability that team $r$ is awarded the point. Ideal (dashed) and the global average (solid) patterns are also shown. }
\end{center}
\label{fig:all:competitions}
\end{figure*}

\vspace{4mm}\noindent \textbf{Competition data.}
Our data are drawn from the popular online game \textit{Halo:\hspace{0.8mm}Reach}, and span nearly 1 billion scoring events across roughly 10 million diversely structured team competitions. These competitions are divided into 125 types according to 35 structural features defining the spatial environmental, competition rules, resource quality, and whether teams had roughly equal skill (see \SI).

\textit{Halo} competitions are a kind of real-time virtual combat. Human players guide their avatars through an arena containing complex terrain, coordinate actions with teammates through visual and audio signals, and encounter opponents. A scoring event occurs when one avatar eliminates another, and this event increments the former's team score. After a short delay, the latter is returned to the competition at another arena location. Competitions end either when a fixed time limit is reached (typically 10 minutes) or when one team's score reaches some threshold (typically 50).

Only individual player skill persists across competitions. Temporary resources, whose control may yield a competitive advantage, are acquirable within a competition, e.g., highly defensible positions, high quality avatar items, and tactical information. Team membership is also temporary, being substantially randomized across competitions by the online system. These features make \textit{Halo} competitions well suited for investigating the impact of structural heterogeneities on competition dynamics. Unlike professional sports, whose team memberships persist across competitions and which exhibit little structural variation, each \textit{Halo} competition is roughly independent of the next, which mitigates confounding effects in characterizing the importance of structural variations.

From the scoring events within a given type of competition, we estimate both model parameters and the outcome predictability (see \SI). This produces a set of coordinates $(\hat{\lambda}_{0},\hat{\alpha},\hat{\beta},\hat{\rho})$ and provides a compact and interpretable summary of that competition type's scoring dynamics and variability. Letting $\vec{\eta}$ denote the structural features of a given competition type, explaining variation across the estimated coordinates from variation in $\vec{\eta}$ will reveal the impact of structural features on competition dynamics, if any.

The determinants of balance $\beta$, which quantifies the strength and distribution of competitive advantages, are of particular interest. Players may prefer more balance because it offers a fair chance at winning. Or, they may prefer less balance because it offers greater reward for the risk. In  these competitions, more balance moderately correlates with a lower probability that at least one player will prematurely leave the field of play ($r^{2}=0.43$, see \SI), a typically voluntary action. Thus, players exhibit a moderate but real preference for more balanced, i.e., more ideal, competitions, whose outcomes are less predictable, whose final score differences are smaller, and whose dynamics are effectively more like a simple stochastic process.

\vspace{4mm}\noindent \textbf{Patterns in Tempo and Score Dynamics.}
We first verify that our generative model effectively captures the true scoring dynamics of these competitions and whether they exhibit patterns similar to those of professional sports.

Across all competition types, we find a consistent three-phase non-stationary pattern in the tempo of scoring events, i.e., the probability of a scoring event as a function of time elapsed or time remaining.  Specifically, we find an early phase of little or uneven activity, a protracted middle phase of slow and steadily increasing activity, and an end phase of either slightly decreased or markedly increased activity (Fig.\ 1
\textit{A}-\textit{C}).

The early- and end-phase patterns are caused by boundary effects in the length of competition, and these are also observed in professional sports~\cite{gabel:redner:2012}. Early in a competition, players require some time to move from their initial positions to their first scoring opportunities, which suppresses the tempo of events relative to the ideal case. Although the shape of this early phase varies moderately by competition type (Fig.\ 1
\textit{A}), after 20--30 seconds these variations largely disappear and the tempo transitions into the more stable middle phase.

Similarly, near a competition's end, the impending cessation of scoring opportunities encourages different strategic choices~\cite{myerson:1997} than in the early or middle phases. Here, we observe either slightly decreased or strongly increased tempo (Fig.\ 1
\textit{C}), depending on whether the competition type's particular rules provide an incentive for risk taking in the final seconds. When the incentive is present, the tempo increases dramatically just before the competition ends, as players take greater risks for the win---a pattern also observed in professional sports~\cite{gabel:redner:2012,thomas2007inter}. When the incentive is absent, players instead adopt defensive positions to deny the opposing team additional points, leading to decreased scoring rates---a pattern not typically observed in sports.

In contrast, the middle phase's tempo exhibits a roughly linear increase over time (Fig.\ 1
\textit{B}), which agrees with our generative model for event timing. To estimate our tempo model parameters, we eliminate the boundary effects by focusing on events in this phase alone (see \SI). Across competition types, both the base tempo and the acceleration vary widely: base rates can vary by up to a factor of two and we observe increases in tempo of 5--20\% over the phase. Within-competition learning is one likely explanation for this increase~\cite{thompson:2010}. Through trial and error, teams may learn how and where to produce scoring events, which progressively reduces the time spent searching for new scoring opportunities.

To understand the variation in the accumulation of points, we examine the distributions of scoring biases across competition types. For a particular competition, the scoring bias is estimated as the fraction of points held by an arbitrarily labeled team $r$. We find that all competition types exhibit moderately non-ideal variations in scoring biases (Fig.\ 1
\textit{D}), i.e., they are consistently dispersed from the ideal case of $c=1/2$.  As with the competition tempo in the middle phase, the degree of dispersion varies substantially across competition types, suggesting a significant role for structural variables.

As a further test of our generative model's quality for these competitions, we estimate $\lambda_{0}$, $\alpha$ and $\beta$ from the entire data set, draw many synthetic competitions from the fitted model, and consider whether the simulated scoring dynamics are similar to those in the empirical data. The results indicate that the simulated competitions match the observed sequences on multiple scoring and timing statistics unrelated to parameter estimation (see \SI). This quantitative agreement indicates that our model successfully captures the important dynamical features of our competitions.

\vspace{4mm}\noindent \textbf{How structure shapes dynamics.}
We now investigate four specific types of structure and their impact on the estimated competition dynamics. These analysis are intended to shed light on how specific structures may shape dynamics, and will aid the interpretation of our systematic analysis below.
\\

\noindent \textit{Team skill differences.}
When assigning individuals to a new competition instance, the online system uses a matchmaking algorithm to substantially randomize team composition. This algorithm operates in two modes. For players who have completed a moderate number of competitions, it adjusts team memberships so that teams have roughly equal total skill. These estimates are derived from a Bayesian generalization of the popular Elo rating system of individual player skill~\cite{herbrich:minka:graepel:2007}. Otherwise, teams are assembled without regard to player skill. We examine the differences in our model parameters for all competitions constructed under each of the two modes.

Differences in skill have a substantial impact on competition balance, as we might expect. However, they have little impact on competition tempo (Table~\ref{table:vignettes}, Fig.\ S4\textit{A}). When teams have roughly equal skill, scoring is more balanced than when the equal-skill control is absent ($\beta=45.9\pm0.35$ versus $20.9\pm0.22$). This difference implies that well-matched teams produce substantially more ideal competitions, have smaller competitive advantages, and exhibit overall dynamics that are closer to those produced by a fair coin. In effect, reducing the difference in team skill serves to amplify the importance of chance events, i.e., accidents and miscalculations.
\\

\begin{table}[t!]
\caption{
Estimated tempo and scoring parameters for four dimensions of competition variation, illustrating a substantial impact of structure on dynamics. Values in parentheses give the bootstrap uncertainty.}
\begin{center}
\begin{tabular}{ll|ccc}
& & balance  & base tempo & acceleration \\
feature\hspace{-1mm} & variation & $\hat{\beta}$ & $\hat{\lambda}_{0}$ ($\times10^{-3}$) & $\hat{\alpha}$ ($\times10^{-5}$) \\
\hline
\multirow{2}{*}{skill}\hspace{-1mm} & equal & 45.9(0.35)  & 166(0.1) & $7.09(0.09)$ \\  
 & unequal & 20.9(0.22)  & 160(0.1) & $7.18(0.02)$ \\  
\hline
\multirow{2}{*}{environment}\hspace{-1mm} & neutral & 47.9(1.20)  & 169(0.4) & $9.09(0.22)$ \\ 
 & irregular & 23.9(0.67)  & 147(0.3) & $7.49(0.21)$ \\ 
\hline
\multirow{2}{*}{scoring}\hspace{-1mm} & standard & 41.7(0.36) &  185(0.2) & $8.45(0.16)$ \\  
 & easy & 30.3(0.71) & 158(1.1) & $9.16(0.64)$ \\ 
\hline
\multirow{2}{*}{resources}\hspace{-1mm} & versatile & 20.2(0.52) & 153(0.2) & $7.08(0.13)$ \\ 
 & limited & 41.7(1.04) & 166(0.3) & $8.49(0.21)$ \\  
\hline \hline
all\hspace{-1mm} & -- & 29.5(0.21) & 163(0.1) &  7.13(0.05)
\end{tabular}
\end{center}
\label{table:vignettes}
\end{table}

\noindent \textit{Physical environment.} 
The arenas for these competitions are typically complex virtual terrains, and may contain large outdoor spaces, complicated indoor corridor systems, buildings with multiple levels, defensible positions, high ground, etc. We compare model parameters for all competitions taking place within two structurally distinct environments: one is largely neutral, exhibiting strong spatial symmetries and few features like defensible locations that might offer tactical advantage, while the other is strongly irregular, with an asymmetric and strongly vertical spatial structure, truncated sight lines, and at least one defensible location.

Overall, the more symmetric environment produces substantially more balanced outcomes and higher scoring rates than the irregular one. In fact, the observed difference in balance parameters is roughly as large as the difference induced by the equal-skill criterion (Table~\ref{table:vignettes}, Fig.\ S4\textit{B}). This suggests that increasing the homogeneity of the competitive environment, e.g., introducing symmetries, removing defensible positions, etc., serves to limit environmental opportunities for competitive advantage. Much like eliminating differences in skill, simpler environments effectively amplify the importance of chance events, making competition scoring more ideal.
\\

\begin{figure}[t!]
\begin{center}
\vspace{-7mm}
\includegraphics[scale=0.38]{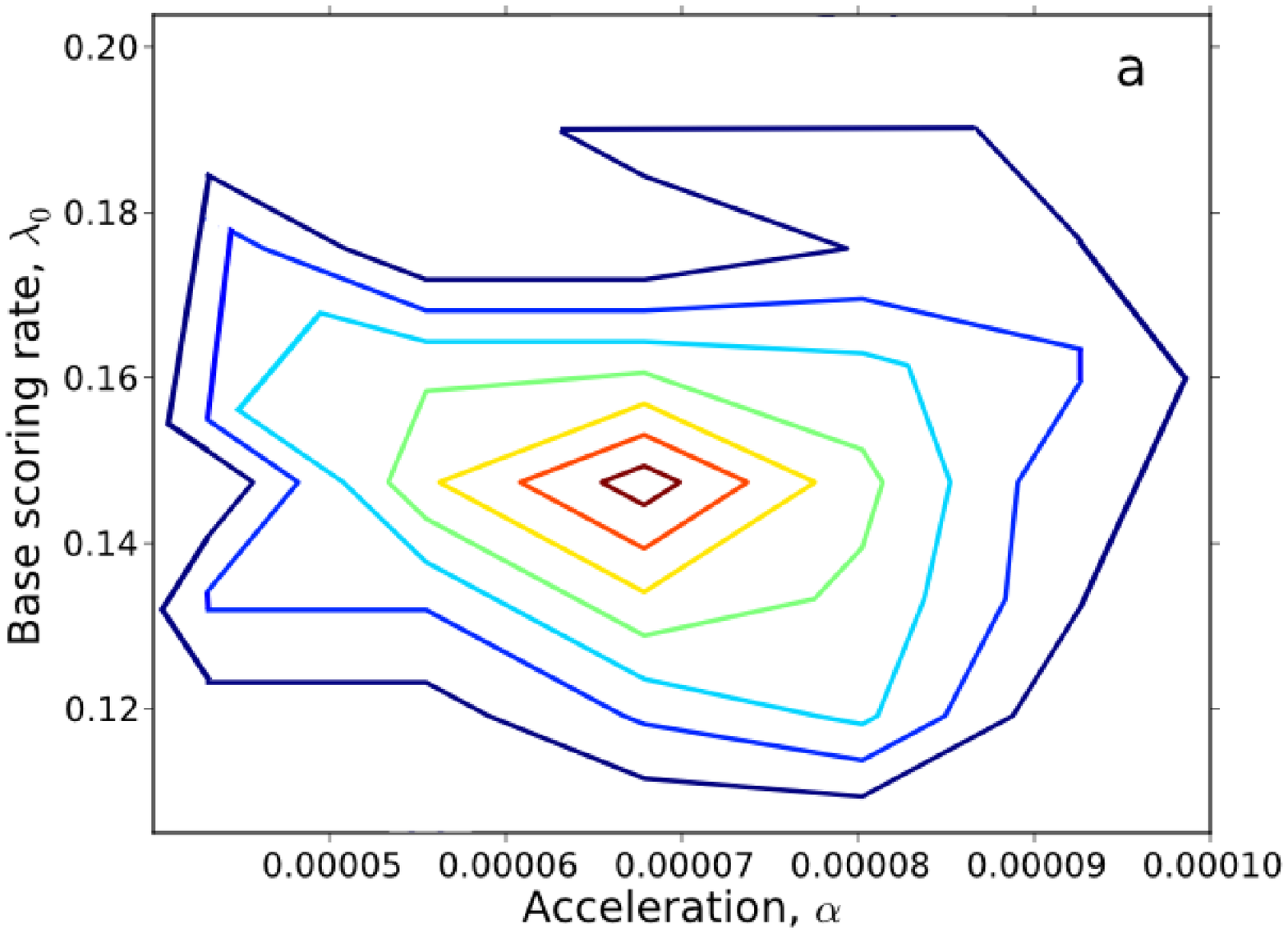}
\includegraphics[scale=0.38]{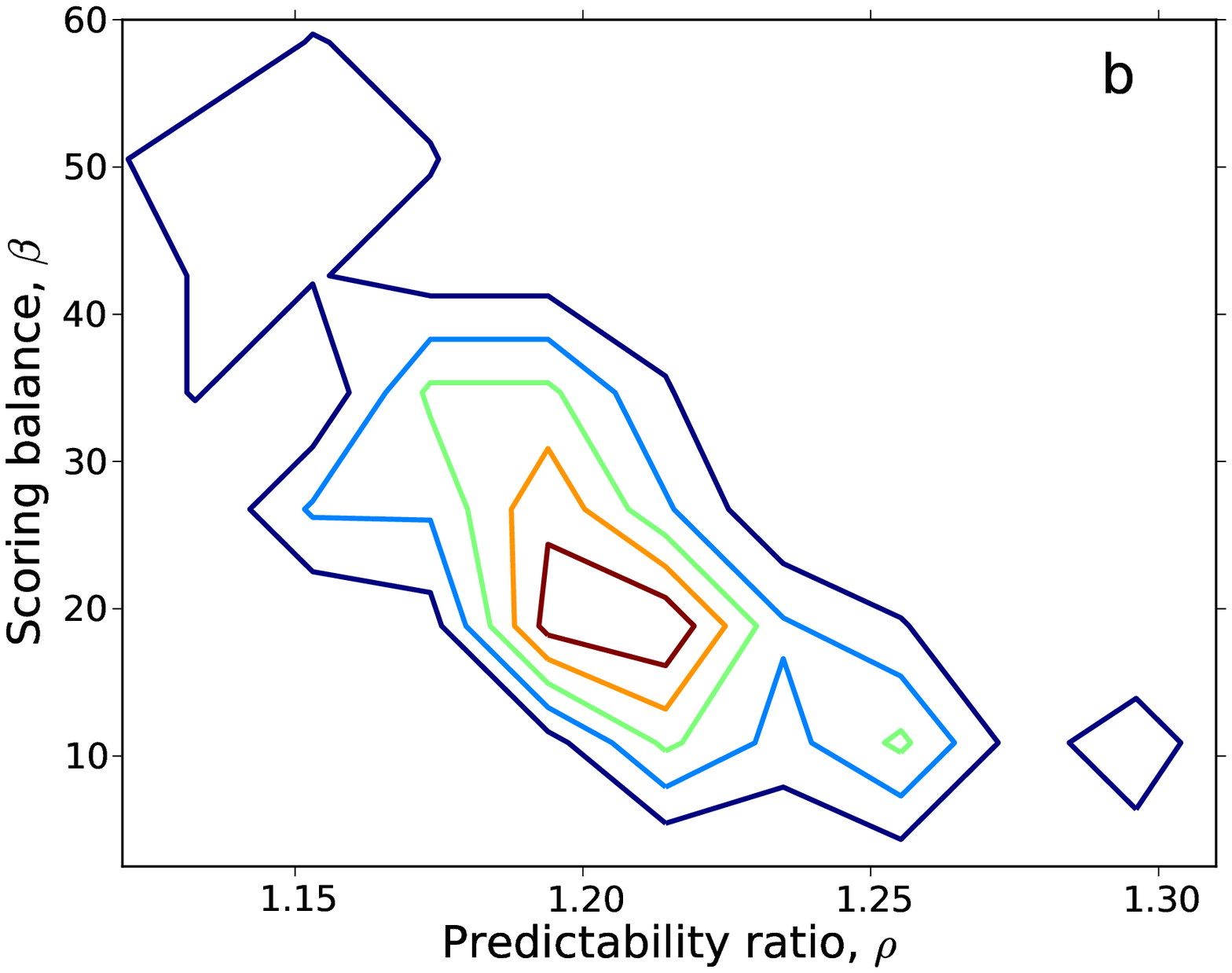}
\caption{Equally spaced quantiles of joint distributions across 125 competition types of (\textit{A}) base scoring rate $\lambda_{0}$ and acceleration $\alpha$, and (\textit{B}) outcome balance $\beta$ and predictability ratio $\rho$. For event timing parameters, we observe little statistical correlation, while greater balance is strongly correlated with lower outcome predictability.}
\end{center}
\label{fig:parameter:maps}
\end{figure}

\noindent \textit{Scoring difficulty.} 
Few studies have examined the difference in competition dynamics caused by variations in the rules of the competition. Our data include several variations of this kind, and we examine one particular variant to shed light on how small changes in rules may impact competition dynamics. A popular group of competition types alters the standard scoring rules by reducing the threshold required to eliminate an opposing avatar and by slightly limiting each player's visual field. These changes make scoring opportunities easier to exploit, and we compare the estimated model parameters for all competitions of the standard and easy scoring types.

Lowering the threshold for scoring has a substantial impact on competition dynamics (Table~\ref{table:vignettes}, Fig.\ S4\textit{C}), with easier scoring rules producing less balanced outcomes. The size of this difference is nearly half as large as the impact of the equal-skill criterion. Additionally, the lower threshold decreases the base scoring rate by 15\% but increases the acceleration  by roughly 8\% over those of standard competitions. The implication is that lowering the barrier to scoring skews the playing field, allowing skilled players to exploit either their skill-based competitive advantage or other structurally-derived advantages.
\\

\noindent \textit{Resource quality.} 
Each competition has a fixed a set of acquirable resources, which players use to score points. Each resource belongs to one of two classes, which we label ``versatile'' and ``limited.'' Versatile resources are generally of higher quality and are more effective for scoring points. When resources of both classes are present in a competition, 80\% of scoring events are associated with the versatile class, illustrating a strong player preference for more effective tools. To clearly separate their effects, we examine competitions with either only versatile- or only limited-class resources.

Limited-class competitions produced moderately higher base and acceleration rates than versatile-class competitions, indicating an overall faster tempo. Furthermore, competitions with only limited-class resources produce substantially more balanced scoring outcomes ($\beta=41.7\pm1.04$ versus $20.2\pm0.52$; Table~\ref{table:vignettes}, Fig.\ S4\textit{D}), a difference as large as that of the equal-skill criterion. Just as environmental structures can be exploited for competitive advantages, differences in the quality of acquirable resources also represent exploitable structural heterogeneities, and limiting such variations can effectively level a playing field to produce more ideal dynamics.

\vspace{4mm}\noindent \textbf{Structural determinants of competitive dynamics.}
Each competition type defines a point on a $(\lambda_{0},\alpha,\beta,\rho)$-manifold, and the distribution of these points describes the observed variability in competition dynamics. We now consider the degree to which a competition's position in this coordinate space is predictable from its structural features alone.

The joint distribution of the model timing parameters $\lambda_{0}$ and $\alpha$ is broadly distributed and shows little internal structure (Fig.\ 2\textit{A}). 
The typical scoring base rate is roughly one event per 7.5 seconds, with variations of 2.5s in either direction. Additionally, nearly all competitions types show modest acceleration rates, with an increase of 10--12\% over the middle-phase of competition being common. The estimated balance parameters $\beta$ are also broadly distributed, indicating a wide range of competitive advantages. The typical competition type has $\beta$ between 20 and 30, but some have values as large as 50 or as small as 10 (Fig.\ 2\textit{B}).
We also observe a strong negative correlation between scoring balance $\beta$ and the predictability $\rho$ of a competition's winner, although with some variation, particularly in the low-$\beta$ regime.

\begin{table}[t!]
\caption{
Ordered multivariate regression coefficients, with uncertainty, for predicting $\hat{\beta}$, $\hat{\lambda}_0$ and $\hat{\alpha}$ of standard-type competitions from structural features alone, and the corresponding fraction of variance explained $r^{2}$. Here, we show only the statistically significant features ($p\ll0.001$, $t$-test); Table~S6 provides the full results.}
\begin{center}
\begin{tabular}{c|lcc|c}
 & structural feature & $\hat{\theta}$ & std. error & $r^2$ \\ \hline
\multirow{11}{*}{$\lambda_{0}$} & E5\hspace{3.2mm}indoor terrain & ~0.082 & 0.008  & \multirow{11}{*}{0.96} \\
& E11\hspace{1.8mm}large arena & ~0.059 & 0.003 &   \\
& E1\hspace{3.2mm}open terrain & ~0.045 & 0.009 &     \\
& E3\hspace{3.2mm}circular terrain & ~0.029 & 0.006 &    \\
& E9\hspace{3.2mm}outdoor terrain & ~0.023 & 0.001 &    \\
& S1\hspace{3.5mm}equally skilled teams & ~0.005 &  0.001 &   \\
& R1\hspace{3.3mm}short \& medium range\hspace{-4mm} & -0.021 & 0.008 &   \\
& R4\hspace{3.3mm}short \& long range & -0.030 & 0.008 &  \\
& R15\hspace{1.9mm}high-quality resources\hspace{-3mm} & -0.032 & 0.006   & \\
& E2\hspace{3.2mm}vertical environment & -0.081 & 0.006   & \\
& E7\hspace{3.2mm}high ground & -0.081 & 0.005   & \\ \hline \hline
\multirow{2}{*}{$\alpha$} & R12\hspace{1.9mm}long range & -1.9x$10^{-5}$  & 8.1x$10^{-6}$  & \multirow{2}{*}{0.65} \\ 
& S1\hspace{3.5mm}equally skilled teams & -2.9x$10^{-6}$ & 1.7x$10^{-6}$  & \\ \hline \hline
\multirow{7}{*}{$\log \beta$} & E5\hspace{3.2mm}indoor terrain & ~1.849  &  0.320 & \multirow{7}{*}{0.93} \\
& E1\hspace{3.2mm}open terrain & ~1.391  &  0.371 &   \\
& E11\hspace{1.8mm}large arena & ~1.123  &  0.141  &  \\
& S1\hspace{3.5mm}equally skilled teams & ~0.822 & 0.034 &   \\
& E9\hspace{3.2mm}outdoor terrain & ~0.481 & 0.076  & \\
& E6\hspace{3.2mm}defensible positions & -0.813 & 0.150 & \\
& E2\hspace{3.2mm}vertical environment & -1.645 &   0.336   & \\
& E7\hspace{3.2mm}high ground & -2.126  &  0.224   & \\ \hline \hline
\multirow{6}{*}{$\rho$} & E7\hspace{3.2mm}high ground & ~0.138 &  0.022 & \multirow{6}{*}{0.89} \\ 
& E2\hspace{3.2mm}vertical environment & ~0.123  & 0.024 & \\
& E6\hspace{3.2mm}defensible positions & ~0.061 &  0.014 & \\
& E9\hspace{3.2mm}outdoor terrain & -0.036 &  0.007 & \\ 
& S1\hspace{3.5mm}equally skilled teams & -0.055 &  0.003 & \\
& E11\hspace{1.8mm}large arena & -0.089 &  0.013 & \\
\end{tabular}
\end{center}
\label{table:lin:reg}
\end{table}%

\vspace{4mm}\noindent \textbf{Predicting dynamics from structure.}
The extent to which a competition's dynamical variables $(\lambda_{0},\alpha,\beta,\rho)$ are predictable from its structural variables $\vec{\eta}$ provides a direct measure of how competition structure shapes dynamics. Thirty-five structural features, divided into resources (R), environment (E), team skill (S), and rules (P) categories, were used to identify 125 distinct types of competition. Regressing these structural features onto the estimated model parameters quantifies the overall predictability of dynamics from structure. The relative importance of these features provides additional insight.

Overall, competition dynamics are highly predictable from structure alone 
(Table~\ref{table:lin:reg}), 
with structural variables explaining 65--96\% of the variance in individual dynamical parameters. Because the coverage across our feature space is sparse, we performed three additional tests to determine the robustness of our results. Both multiple and stepwise regressions produce models of nearly equal quality and assign features nearly the same relative importances. Randomizing the association of structural and dynamical variables yields non-significant correlations (see \SI), indicating our results are reliable.

Competition structure has the largest impact on base rate $\lambda_{0}$ ($r^{2}=0.96$), and features describing neutral or homogeneous environments play the dominant role in setting its value. The base scoring rate is effectively determined by the ``encounter rate'' between scoring opportunities and competitors. In these competitions, an encounter requires two individuals to locate and engage each other; thus, small, neutral environments generate these encounters more often than large, irregular ones. Competitions between equally-skilled teams exhibit a higher encounter rate, but only marginally, as the skill coefficient is four time smaller in absolute value than any other statistically significant feature.

The change in scoring rate $\alpha$ is moderately well predicted by structure ($r^{2}=0.65$), and competitions with resources that operate across long ranges and with well-matched teams exhibit less acceleration over the middle phase. These resources make it easier to locate and exploit the next scoring opportunity, thus mitigating the difficulty of searching for new opportunities within large or irregular environments. Similarly, skilled competitors tend to have prior experience with the location of resources and strategic environmental structures, improving their search efficiency and lowering $\alpha$.

The scoring balance $\beta$, which measures the strength of the associated competitive advantages, is highly predictable from structure ($r^2=0.93$, regression on $\log \beta$), as is the relative predictability $\rho$ of the winning team ($r^{2}=0.89$). Having well-matched teams, however, is only moderately important for increasing balance, and well balanced scoring is typically derived from large, neutral environments, a situation similar to professional team sports with their level playing fields. However, the single feature that produces the most balanced competitions, by a factor of two, is indoor terrain, i.e., rooms and corridors. This particular form spatial heterogeneity may effectively handicap all competitors by limiting their spatial awareness, thus mitigating other competitive advantages, including those derived from greater skill or more versatile resources, thereby making scoring opportunities and outcomes less predictable and more ideal.

In contrast, the most imbalanced and predictable competitions are those with controllable or strategically valuable environmental features like high ground or defensible positions. For setting the values of $\beta$ and $\rho$, such features are at least as important, but opposite in sign, to having teams of equal skill. These strategically important environmental features can thus effectively upset the competitive balance produced by well-matched teams by providing one team with a sustained competitive advantage throughout the competition.

Surprisingly, variation in rules, including reduced spatial awareness, weakened defensive capabilities, or a lower threshold for scoring, were not statistically significant predictors. None of these features produced a measurable impact on the tempo or balance of scoring within competitions, once the effects of other features were taken into account.

\section*{Discussion}
Although professional sports are often considered models of team competition~\cite{gabel:redner:2012,thomas2007inter,heuer2012does,sire:redner:2009,ben2005most}, their limited structural variation provides few opportunities for understanding how competition structure can shape competition dynamics. Our results shed new light on these and other fundamental questions about human social dynamics and competition.

In particular, heterogeneities in the spatial environment, available resources, competition rules, and team skill exert a strong influence on the balance and tempo of scoring within a competition. For the virtual team-combat simulation studied here, spatial structure plays the most important role in producing competitive advantages, with skill and resource differences assuming supporting roles. It is thus not a superficial analogy to say that like business firms leveraging heterogeneous and scarce resources for sustained competitive advantage in a marketplace~\cite{barney:1991}, teams in \textit{Halo} leverage environmental and resource heterogeneities, like high ground and defensible positions, toward the same ends.

But unlike the pattern of either business firms or professional sports teams, some heterogeneities---in the case of \textit{Halo}, significant indoor terrain---can effectively neutralize competitive advantages normally derived from exploitable structural features. When these ``leveling'' features are present, scoring outcomes are substantially more balanced than when they are absent, and this leveling effect is stronger than the one produced by having equally skilled teams. Although the precise mechanisms of these leveling effects remain unknown, their existence implies that competitive advantages are derived from specific mechanisms whose effects can be neutralized by other mechanisms. A better understanding of these mechanisms could be derived from controlled experiments with level design, and may facilitate the design of inhomogeneous competitive environments that nevertheless exhibit the balanced dynamics that homogeneous environments produce.

Otherwise, the most ideal competitions do indeed occur in large neutral spaces between well-matched teams. It is thus no accident that professional team sports are often played in precisely this type of environment: absent spatial or resource heterogeneity, competition between skilled teams is significantly more ideal. Counterintuitively, the more ideal a competition, the more effectively it may be described as a purely random process, not despite but in fact because of the significant strategic and tactical effort behind individual events. That is, the more ideal a competition, the greater the role of chance events like miscalculations and accidents in determining the outcome. We note, however, that replacing the underlying competition mechanics by actual coin flipping seems unlikely to produce the same level or type of engagement among players and spectators.

The three-phase pattern in the tempo of events in \textit{Halo} competitions is strikingly similar to the pattern observed in professional team sports~\cite{gabel:redner:2012}. Yet the underlying structures of most professional sports and a team combat simulation could hardly be more different. In the former, goals have fixed locations, the environment and within-competition resources are homogeneous, and teams are highly trained and persistent. In the latter, goals are highly mobile, the environment and within-competition resources are heterogeneous, and teams are largely non-persistent. The existence of a common dynamical pattern despite such differences suggests that it may be a universal feature of team competitions. The elucidation of its origin is an important open question.

Finally, we omitted explicit roles for within-team variables like team composition~\cite{guimera:etal:2005}, coordination~\cite{mason:clauset:2012}, and player characteristics. Their impact is implicit within the estimated model parameters, whose variation is well explained by structural variables alone. This particular result is likely supported by the substantial randomization in team membership across \textit{Halo} competitions, which serves to mitigate any significant differences in team composition. Player and team characteristics likely play a more significant role in determining the dynamics in competitions with persistent teams or homogeneous environments, as in professional sports. A broad study of within-competition dynamics across fundamentally different types of competition may shed complementary light on the origin of competitive advantages, the mechanisms by which specific features promote or discourage balanced outcomes, and the fundamental laws of competitive dynamics, if any.

\begin{acknowledgments}
We thank Cris Moore, Sid Redner, Tim Brown, Christopher Aicher, Dan Larremore, Abigail Jacobs, Winter Mason, FearfulSpoon, Stilted Fox, AngryKnife, jizackalyn and TOURIST1224 for helpful conversations, and Chris Schenk for help developing the data acquisition system. This work was supported in part by the James S.~McDonnell Foundation.
\end{acknowledgments}

\bibliography{competition_refs}

\begin{thebibliography}{27}%
\makeatletter
\providecommand \@ifxundefined [1]{%
 \@ifx{#1\undefined}
}%
\providecommand \@ifnum [1]{%
 \ifnum #1\expandafter \@firstoftwo
 \else \expandafter \@secondoftwo
 \fi
}%
\providecommand \@ifx [1]{%
 \ifx #1\expandafter \@firstoftwo
 \else \expandafter \@secondoftwo
 \fi
}%
\providecommand \natexlab [1]{#1}%
\providecommand \enquote  [1]{``#1''}%
\providecommand \bibnamefont  [1]{#1}%
\providecommand \bibfnamefont [1]{#1}%
\providecommand \citenamefont [1]{#1}%
\providecommand \href@noop [0]{\@secondoftwo}%
\providecommand \href [0]{\begingroup \@sanitize@url \@href}%
\providecommand \@href[1]{\@@startlink{#1}\@@href}%
\providecommand \@@href[1]{\endgroup#1\@@endlink}%
\providecommand \@sanitize@url [0]{\catcode `\\12\catcode `\$12\catcode
  `\&12\catcode `\#12\catcode `\^12\catcode `\_12\catcode `\%12\relax}%
\providecommand \@@startlink[1]{}%
\providecommand \@@endlink[0]{}%
\providecommand \url  [0]{\begingroup\@sanitize@url \@url }%
\providecommand \@url [1]{\endgroup\@href {#1}{\urlprefix }}%
\providecommand \urlprefix  [0]{URL }%
\providecommand \Eprint [0]{\href }%
\providecommand \doibase [0]{http://dx.doi.org/}%
\providecommand \selectlanguage [0]{\@gobble}%
\providecommand \bibinfo  [0]{\@secondoftwo}%
\providecommand \bibfield  [0]{\@secondoftwo}%
\providecommand \translation [1]{[#1]}%
\providecommand \BibitemOpen [0]{}%
\providecommand \bibitemStop [0]{}%
\providecommand \bibitemNoStop [0]{.\EOS\space}%
\providecommand \EOS [0]{\spacefactor3000\relax}%
\providecommand \BibitemShut  [1]{\csname bibitem#1\endcsname}%
\let\auto@bib@innerbib\@empty
\bibitem [{\citenamefont {Reed}\ and\ \citenamefont
  {Hughes}(2006)}]{reed:hughes:2006}%
  \BibitemOpen
  \bibfield  {author} {\bibinfo {author} {\bibfnamefont {D.}~\bibnamefont
  {Reed}}\ and\ \bibinfo {author} {\bibfnamefont {M.}~\bibnamefont {Hughes}},\
  }\href@noop {} {\bibfield  {journal} {\bibinfo  {journal} {International
  Journal of Performance Analysis in Sport}\ }\textbf {\bibinfo {volume} {6}},\
  \bibinfo {pages} {114} (\bibinfo {year} {2006})}\BibitemShut {NoStop}%
\bibitem [{\citenamefont
  {J\'er\.{o}me~Bourbousson}(2012)}]{bourbousson:seve:mcgarry:2012}%
  \BibitemOpen
  \bibfield  {author} {\bibinfo {author} {\bibfnamefont {C.~S. . T.~M.}\
  \bibnamefont {J\'er\.{o}me~Bourbousson}},\ }\href@noop {} {\bibfield
  {journal} {\bibinfo  {journal} {J. Sports Sciences}\ }\textbf {\bibinfo
  {volume} {28}},\ \bibinfo {pages} {349} (\bibinfo {year} {2012})}\BibitemShut
  {NoStop}%
\bibitem [{\citenamefont {Galla}\ and\ \citenamefont
  {Farmer}(2013)}]{galla:farmer:2013}%
  \BibitemOpen
  \bibfield  {author} {\bibinfo {author} {\bibfnamefont {T.}~\bibnamefont
  {Galla}}\ and\ \bibinfo {author} {\bibfnamefont {J.~D.}\ \bibnamefont
  {Farmer}},\ }\href@noop {} {\bibfield  {journal} {\bibinfo  {journal} {Proc.\
  Natl.\ Acad.\ Sci.\ (USA)}\ }\textbf {\bibinfo {volume} {110}},\ \bibinfo
  {pages} {1232} (\bibinfo {year} {2013})}\BibitemShut {NoStop}%
\bibitem [{\citenamefont {Myerson}(1997)}]{myerson:1997}%
  \BibitemOpen
  \bibfield  {author} {\bibinfo {author} {\bibfnamefont {R.~B.}\ \bibnamefont
  {Myerson}},\ }\href@noop {} {\emph {\bibinfo {title} {Game Theory: Analysis
  of Conflict}}}\ (\bibinfo  {publisher} {Harvard University Press},\ \bibinfo
  {address} {Cambridge MA},\ \bibinfo {year} {1997})\BibitemShut {NoStop}%
\bibitem [{\citenamefont {Denrell}(2004)}]{denrell:2004}%
  \BibitemOpen
  \bibfield  {author} {\bibinfo {author} {\bibfnamefont {J.}~\bibnamefont
  {Denrell}},\ }\href@noop {} {\bibfield  {journal} {\bibinfo  {journal}
  {Management Science}\ }\textbf {\bibinfo {volume} {50}},\ \bibinfo {pages}
  {922} (\bibinfo {year} {2004})}\BibitemShut {NoStop}%
\bibitem [{\citenamefont {Herbrich}\ \emph {et~al.}(2007)\citenamefont
  {Herbrich}, \citenamefont {Minka},\ and\ \citenamefont
  {Graepel}}]{herbrich:minka:graepel:2007}%
  \BibitemOpen
  \bibfield  {author} {\bibinfo {author} {\bibfnamefont {R.}~\bibnamefont
  {Herbrich}}, \bibinfo {author} {\bibfnamefont {T.}~\bibnamefont {Minka}}, \
  and\ \bibinfo {author} {\bibfnamefont {T.}~\bibnamefont {Graepel}},\
  }\href@noop {} {\bibfield  {journal} {\bibinfo  {journal} {Advances in Neural
  Information Processing Systems}\ }\textbf {\bibinfo {volume} {20}},\ \bibinfo
  {pages} {569} (\bibinfo {year} {2007})}\BibitemShut {NoStop}%
\bibitem [{\citenamefont {Barney}(1991)}]{barney:1991}%
  \BibitemOpen
  \bibfield  {author} {\bibinfo {author} {\bibfnamefont {J.}~\bibnamefont
  {Barney}},\ }\href@noop {} {\bibfield  {journal} {\bibinfo  {journal} {J.
  Management}\ }\textbf {\bibinfo {volume} {17}},\ \bibinfo {pages} {99}
  (\bibinfo {year} {1991})}\BibitemShut {NoStop}%
\bibitem [{\citenamefont {Ben-Naim}\ \emph {et~al.}(2012)\citenamefont
  {Ben-Naim}, \citenamefont {Hengartner}, \citenamefont {Redner},\ and\
  \citenamefont {Vazquez}}]{bennaim:etal:2012}%
  \BibitemOpen
  \bibfield  {author} {\bibinfo {author} {\bibfnamefont {E.}~\bibnamefont
  {Ben-Naim}}, \bibinfo {author} {\bibfnamefont {N.}~\bibnamefont
  {Hengartner}}, \bibinfo {author} {\bibfnamefont {S.}~\bibnamefont {Redner}},
  \ and\ \bibinfo {author} {\bibfnamefont {F.}~\bibnamefont {Vazquez}},\
  }\href@noop {} {\enquote {\bibinfo {title} {Randomness in competitions},}\ }
  (\bibinfo {year} {2012}),\ \bibinfo {note} {preprint,
  http://arxiv.org/abs/1209.4724}\BibitemShut {NoStop}%
\bibitem [{\citenamefont {Salganik}\ \emph {et~al.}(2006)\citenamefont
  {Salganik}, \citenamefont {Dodds},\ and\ \citenamefont
  {Watts}}]{salganik:etal:2006}%
  \BibitemOpen
  \bibfield  {author} {\bibinfo {author} {\bibfnamefont {M.~J.}\ \bibnamefont
  {Salganik}}, \bibinfo {author} {\bibfnamefont {P.~S.}\ \bibnamefont {Dodds}},
  \ and\ \bibinfo {author} {\bibfnamefont {D.~J.}\ \bibnamefont {Watts}},\
  }\href@noop {} {\bibfield  {journal} {\bibinfo  {journal} {Science}\ }\textbf
  {\bibinfo {volume} {311}},\ \bibinfo {pages} {854} (\bibinfo {year}
  {2006})}\BibitemShut {NoStop}%
\bibitem [{\citenamefont {Denrell}\ and\ \citenamefont
  {Liu}(2012)}]{denrell:liu:2012}%
  \BibitemOpen
  \bibfield  {author} {\bibinfo {author} {\bibfnamefont {J.}~\bibnamefont
  {Denrell}}\ and\ \bibinfo {author} {\bibfnamefont {C.}~\bibnamefont {Liu}},\
  }\href@noop {} {\bibfield  {journal} {\bibinfo  {journal} {Proc.\ Natl.\
  Acad.\ Sci.\ (USA)}\ }\textbf {\bibinfo {volume} {109}},\ \bibinfo {pages}
  {9331} (\bibinfo {year} {2012})}\BibitemShut {NoStop}%
\bibitem [{\citenamefont {Michael}\ and\ \citenamefont
  {Chen}(2005)}]{michael:chen:2005}%
  \BibitemOpen
  \bibfield  {author} {\bibinfo {author} {\bibfnamefont {D.~R.}\ \bibnamefont
  {Michael}}\ and\ \bibinfo {author} {\bibfnamefont {S.~L.}\ \bibnamefont
  {Chen}},\ }\href@noop {} {\emph {\bibinfo {title} {Serious Games: Games That
  Educate, Train, and Inform}}}\ (\bibinfo  {publisher} {Muska and Lipman},\
  \bibinfo {year} {2005})\BibitemShut {NoStop}%
\bibitem [{nis(2007)}]{nisan:etal:2007}%
  \BibitemOpen
  in\ \href@noop {} {\emph {\bibinfo {booktitle} {Algorithmic Game Theory}}},\
  \bibinfo {editor} {edited by\ \bibinfo {editor} {\bibfnamefont
  {N.}~\bibnamefont {Nisan}}, \bibinfo {editor} {\bibfnamefont
  {T.}~\bibnamefont {Roughgarden}}, \bibinfo {editor} {\bibfnamefont
  {E.}~\bibnamefont {Tardos}}, \ and\ \bibinfo {editor} {\bibfnamefont {V.~V.}\
  \bibnamefont {Vazirani}}}\ (\bibinfo  {publisher} {Cambridge University
  Press},\ \bibinfo {year} {2007})\BibitemShut {NoStop}%
\bibitem [{\citenamefont {Boucher}(1985)}]{boucher:1985}%
  \BibitemOpen
  \bibfield  {author} {\bibinfo {author} {\bibfnamefont {D.~H.}\ \bibnamefont
  {Boucher}},\ }in\ \href@noop {} {\emph {\bibinfo {booktitle} {The Biology of
  Mutualism: Ecology and Evolution}}},\ \bibinfo {editor} {edited by\ \bibinfo
  {editor} {\bibfnamefont {D.~H.}\ \bibnamefont {Boucher}}}\ (\bibinfo
  {publisher} {Oxford University Press},\ \bibinfo {year} {1985})\ pp.\
  \bibinfo {pages} {1--27}\BibitemShut {NoStop}%
\bibitem [{\citenamefont {Bowles}(2006)}]{bowles:2006}%
  \BibitemOpen
  \bibfield  {author} {\bibinfo {author} {\bibfnamefont {S.}~\bibnamefont
  {Bowles}},\ }\href@noop {} {\emph {\bibinfo {title} {Microeconomics:
  Behavior, Institutions, and Evolution}}}\ (\bibinfo  {publisher} {Princeton
  University Press},\ \bibinfo {year} {2006})\BibitemShut {NoStop}%
\bibitem [{\citenamefont {{Entertainment Software
  Association}}(2011)}]{esa:2011}%
  \BibitemOpen
  \bibfield  {author} {\bibinfo {author} {\bibnamefont {{Entertainment Software
  Association}}},\ }\href@noop {} {\enquote {\bibinfo {title} {{Essential Facts
  about the Computer and Video Game Industry}},}\ } (\bibinfo {year} {2011}),\
  \bibinfo {note} {http://bit.ly/kLHJ2Q, (access date February,
  2012)}\BibitemShut {NoStop}%
\bibitem [{\citenamefont {Gabel}\ and\ \citenamefont
  {Redner}(2012)}]{gabel:redner:2012}%
  \BibitemOpen
  \bibfield  {author} {\bibinfo {author} {\bibfnamefont {A.}~\bibnamefont
  {Gabel}}\ and\ \bibinfo {author} {\bibfnamefont {S.}~\bibnamefont {Redner}},\
  }\href@noop {} {\bibfield  {journal} {\bibinfo  {journal} {J. Quant. Analysis
  in Sports}\ }\textbf {\bibinfo {volume} {8}},\ \bibinfo {pages} {Manuscript
  1416} (\bibinfo {year} {2012})}\BibitemShut {NoStop}%
\bibitem [{\citenamefont {Szell}\ \emph {et~al.}(2010)\citenamefont {Szell},
  \citenamefont {Lambiotte},\ and\ \citenamefont
  {Thurner}}]{szell:lambiotte:thurner:2010}%
  \BibitemOpen
  \bibfield  {author} {\bibinfo {author} {\bibfnamefont {M.}~\bibnamefont
  {Szell}}, \bibinfo {author} {\bibfnamefont {R.}~\bibnamefont {Lambiotte}}, \
  and\ \bibinfo {author} {\bibfnamefont {S.}~\bibnamefont {Thurner}},\
  }\href@noop {} {\bibfield  {journal} {\bibinfo  {journal} {Proc.\ Natl.\
  Acad.\ Sci.\ (USA)}\ }\textbf {\bibinfo {volume} {107}},\ \bibinfo {pages}
  {13636} (\bibinfo {year} {2010})}\BibitemShut {NoStop}%
\bibitem [{\citenamefont {Brzezniak}\ and\ \citenamefont
  {Zastawniak}(2000)}]{brzezniak:zastawniak:2000}%
  \BibitemOpen
  \bibfield  {author} {\bibinfo {author} {\bibfnamefont {Z.}~\bibnamefont
  {Brzezniak}}\ and\ \bibinfo {author} {\bibfnamefont {T.}~\bibnamefont
  {Zastawniak}},\ }\href@noop {} {\emph {\bibinfo {title} {Basic Stochastic
  Processes}}}\ (\bibinfo  {publisher} {Springer},\ \bibinfo {address}
  {Berlin},\ \bibinfo {year} {2000})\BibitemShut {NoStop}%
\bibitem [{\citenamefont {Thomas}(2007)}]{thomas2007inter}%
  \BibitemOpen
  \bibfield  {author} {\bibinfo {author} {\bibfnamefont {A.}~\bibnamefont
  {Thomas}},\ }\href@noop {} {\bibfield  {journal} {\bibinfo  {journal} {J.
  Quant. Analysis in Sports}\ }\textbf {\bibinfo {volume} {3}} (\bibinfo {year}
  {2007})}\BibitemShut {NoStop}%
\bibitem [{\citenamefont {Heuer}\ and\ \citenamefont
  {Rubner}(2012)}]{heuer2012does}%
  \BibitemOpen
  \bibfield  {author} {\bibinfo {author} {\bibfnamefont {A.}~\bibnamefont
  {Heuer}}\ and\ \bibinfo {author} {\bibfnamefont {O.}~\bibnamefont {Rubner}},\
  }\href@noop {} {\enquote {\bibinfo {title} {How does the past of a soccer
  match influence its future?}}\ } (\bibinfo {year} {2012}),\ \bibinfo {note}
  {preprint, http://arxiv.org/abs/1207.4471}\BibitemShut {NoStop}%
\bibitem [{\citenamefont {Bishop}(2006)}]{bishop:2006}%
  \BibitemOpen
  \bibfield  {author} {\bibinfo {author} {\bibfnamefont {C.~M.}\ \bibnamefont
  {Bishop}},\ }\href@noop {} {\emph {\bibinfo {title} {Pattern Recognition and
  Machine Learning}}}\ (\bibinfo  {publisher} {Springer},\ \bibinfo {year}
  {2006})\BibitemShut {NoStop}%
\bibitem [{\citenamefont {Bradley}(1997)}]{bradley:1997}%
  \BibitemOpen
  \bibfield  {author} {\bibinfo {author} {\bibfnamefont {A.}~\bibnamefont
  {Bradley}},\ }\href@noop {} {\bibfield  {journal} {\bibinfo  {journal}
  {Pattern Recognition}\ }\textbf {\bibinfo {volume} {30}},\ \bibinfo {pages}
  {1145} (\bibinfo {year} {1997})}\BibitemShut {NoStop}%
\bibitem [{\citenamefont {Thompson}(2010)}]{thompson:2010}%
  \BibitemOpen
  \bibfield  {author} {\bibinfo {author} {\bibfnamefont {P.}~\bibnamefont
  {Thompson}},\ }in\ \href@noop {} {\emph {\bibinfo {booktitle} {Handbook of
  Economics of Technical Change}}},\ \bibinfo {editor} {edited by\ \bibinfo
  {editor} {\bibfnamefont {B.}~\bibnamefont {Hall}}\ and\ \bibinfo {editor}
  {\bibfnamefont {N.}~\bibnamefont {Rosenberg}}}\ (\bibinfo  {publisher}
  {Elsevier/North-Holland},\ \bibinfo {year} {2010})\ pp.\ \bibinfo {pages}
  {429--476}\BibitemShut {NoStop}%
\bibitem [{\citenamefont {Sire}\ and\ \citenamefont
  {Redner}(2009)}]{sire:redner:2009}%
  \BibitemOpen
  \bibfield  {author} {\bibinfo {author} {\bibfnamefont {C.}~\bibnamefont
  {Sire}}\ and\ \bibinfo {author} {\bibfnamefont {S.}~\bibnamefont {Redner}},\
  }\href@noop {} {\bibfield  {journal} {\bibinfo  {journal} {Eur.\ Phys.\ J.\
  B}\ }\textbf {\bibinfo {volume} {67}},\ \bibinfo {pages} {473} (\bibinfo
  {year} {2009})}\BibitemShut {NoStop}%
\bibitem [{\citenamefont {Ben-Naim}\ \emph {et~al.}(2007)\citenamefont
  {Ben-Naim}, \citenamefont {Vazquez},\ and\ \citenamefont
  {Redner}}]{ben2005most}%
  \BibitemOpen
  \bibfield  {author} {\bibinfo {author} {\bibfnamefont {E.}~\bibnamefont
  {Ben-Naim}}, \bibinfo {author} {\bibfnamefont {F.}~\bibnamefont {Vazquez}}, \
  and\ \bibinfo {author} {\bibfnamefont {S.}~\bibnamefont {Redner}},\
  }\href@noop {} {\bibfield  {journal} {\bibinfo  {journal} {J. Korean Phys.
  Soc.}\ }\textbf {\bibinfo {volume} {50}},\ \bibinfo {pages} {124} (\bibinfo
  {year} {2007})}\BibitemShut {NoStop}%
\bibitem [{\citenamefont {Guimer\`{a}}\ \emph {et~al.}(2005)\citenamefont
  {Guimer\`{a}}, \citenamefont {Uzzi}, \citenamefont {Spiro},\ and\
  \citenamefont {Amaral}}]{guimera:etal:2005}%
  \BibitemOpen
  \bibfield  {author} {\bibinfo {author} {\bibfnamefont {R.}~\bibnamefont
  {Guimer\`{a}}}, \bibinfo {author} {\bibfnamefont {B.}~\bibnamefont {Uzzi}},
  \bibinfo {author} {\bibfnamefont {J.}~\bibnamefont {Spiro}}, \ and\ \bibinfo
  {author} {\bibfnamefont {L.~A.~N.}\ \bibnamefont {Amaral}},\ }\href@noop {}
  {\bibfield  {journal} {\bibinfo  {journal} {Science}\ }\textbf {\bibinfo
  {volume} {308}},\ \bibinfo {pages} {697} (\bibinfo {year}
  {2005})}\BibitemShut {NoStop}%
\bibitem [{\citenamefont {Mason}\ and\ \citenamefont
  {Clauset}(2013)}]{mason:clauset:2012}%
  \BibitemOpen
  \bibfield  {author} {\bibinfo {author} {\bibfnamefont {W.}~\bibnamefont
  {Mason}}\ and\ \bibinfo {author} {\bibfnamefont {A.}~\bibnamefont {Clauset}}\
  }(\bibinfo {year} {2013})\ \bibinfo {note} {16th ACM Conference on Computer
  Supported Cooperative Work and Social Computing}\BibitemShut {NoStop}%
\end{thebibliography}%


\begin{thebibliography}{8}%
\makeatletter
\providecommand \@ifxundefined [1]{%
 \@ifx{#1\undefined}
}%
\providecommand \@ifnum [1]{%
 \ifnum #1\expandafter \@firstoftwo
 \else \expandafter \@secondoftwo
 \fi
}%
\providecommand \@ifx [1]{%
 \ifx #1\expandafter \@firstoftwo
 \else \expandafter \@secondoftwo
 \fi
}%
\providecommand \natexlab [1]{#1}%
\providecommand \enquote  [1]{``#1''}%
\providecommand \bibnamefont  [1]{#1}%
\providecommand \bibfnamefont [1]{#1}%
\providecommand \citenamefont [1]{#1}%
\providecommand \href@noop [0]{\@secondoftwo}%
\providecommand \href [0]{\begingroup \@sanitize@url \@href}%
\providecommand \@href[1]{\@@startlink{#1}\@@href}%
\providecommand \@@href[1]{\endgroup#1\@@endlink}%
\providecommand \@sanitize@url [0]{\catcode `\\12\catcode `\$12\catcode
  `\&12\catcode `\#12\catcode `\^12\catcode `\_12\catcode `\%12\relax}%
\providecommand \@@startlink[1]{}%
\providecommand \@@endlink[0]{}%
\providecommand \url  [0]{\begingroup\@sanitize@url \@url }%
\providecommand \@url [1]{\endgroup\@href {#1}{\urlprefix }}%
\providecommand \urlprefix  [0]{URL }%
\providecommand \Eprint [0]{\href }%
\providecommand \doibase [0]{http://dx.doi.org/}%
\providecommand \selectlanguage [0]{\@gobble}%
\providecommand \bibinfo  [0]{\@secondoftwo}%
\providecommand \bibfield  [0]{\@secondoftwo}%
\providecommand \translation [1]{[#1]}%
\providecommand \BibitemOpen [0]{}%
\providecommand \bibitemStop [0]{}%
\providecommand \bibitemNoStop [0]{.\EOS\space}%
\providecommand \EOS [0]{\spacefactor3000\relax}%
\providecommand \BibitemShut  [1]{\csname bibitem#1\endcsname}%
\let\auto@bib@innerbib\@empty
\bibitem [{\citenamefont {{Entertainment Software
  Association}}(2011)}]{esa:2011}%
  \BibitemOpen
  \bibfield  {author} {\bibinfo {author} {\bibnamefont {{Entertainment Software
  Association}}},\ }\href@noop {} {\enquote {\bibinfo {title} {{Essential Facts
  about the Computer and Video Game Industry}},}\ } (\bibinfo {year} {2011}),\
  \bibinfo {note} {http://bit.ly/kLHJ2Q, (access date February,
  2012)}\BibitemShut {NoStop}%
\bibitem [{\citenamefont {Mason}\ and\ \citenamefont
  {Clauset}(2013)}]{mason:clauset:2012}%
  \BibitemOpen
  \bibfield  {author} {\bibinfo {author} {\bibfnamefont {W.}~\bibnamefont
  {Mason}}\ and\ \bibinfo {author} {\bibfnamefont {A.}~\bibnamefont {Clauset}}\
  }(\bibinfo {year} {2013})\ \bibinfo {note} {16th ACM Conference on Computer
  Supported Cooperative Work and Social Computing}\BibitemShut {NoStop}%
\bibitem [{\citenamefont {Herbrich}\ \emph {et~al.}(2007)\citenamefont
  {Herbrich}, \citenamefont {Minka},\ and\ \citenamefont
  {Graepel}}]{herbrich:minka:graepel:2007}%
  \BibitemOpen
  \bibfield  {author} {\bibinfo {author} {\bibfnamefont {R.}~\bibnamefont
  {Herbrich}}, \bibinfo {author} {\bibfnamefont {T.}~\bibnamefont {Minka}}, \
  and\ \bibinfo {author} {\bibfnamefont {T.}~\bibnamefont {Graepel}},\
  }\href@noop {} {\bibfield  {journal} {\bibinfo  {journal} {Advances in Neural
  Information Processing Systems}\ }\textbf {\bibinfo {volume} {20}},\ \bibinfo
  {pages} {569} (\bibinfo {year} {2007})}\BibitemShut {NoStop}%
\bibitem [{\citenamefont {Ruef}\ \emph {et~al.}(2003)\citenamefont {Ruef},
  \citenamefont {Aldrich},\ and\ \citenamefont
  {Carter}}]{ruef:aldrich:carter:2003}%
  \BibitemOpen
  \bibfield  {author} {\bibinfo {author} {\bibfnamefont {M.}~\bibnamefont
  {Ruef}}, \bibinfo {author} {\bibfnamefont {H.~E.}\ \bibnamefont {Aldrich}}, \
  and\ \bibinfo {author} {\bibfnamefont {N.~M.}\ \bibnamefont {Carter}},\
  }\href@noop {} {\bibfield  {journal} {\bibinfo  {journal} {American
  Sociological Review}\ }\textbf {\bibinfo {volume} {68}},\ \bibinfo {pages}
  {195} (\bibinfo {year} {2003})}\BibitemShut {NoStop}%
\bibitem [{\citenamefont {Baldwin}\ \emph {et~al.}(1997)\citenamefont
  {Baldwin}, \citenamefont {Bedell},\ and\ \citenamefont
  {Johnson}}]{baldwin:etal:1997}%
  \BibitemOpen
  \bibfield  {author} {\bibinfo {author} {\bibfnamefont {T.~T.}\ \bibnamefont
  {Baldwin}}, \bibinfo {author} {\bibfnamefont {M.~D.}\ \bibnamefont {Bedell}},
  \ and\ \bibinfo {author} {\bibfnamefont {J.~L.}\ \bibnamefont {Johnson}},\
  }\href@noop {} {\bibfield  {journal} {\bibinfo  {journal} {Acad.\ of Manag.\
  Journal}\ }\textbf {\bibinfo {volume} {40}},\ \bibinfo {pages} {1369}
  (\bibinfo {year} {1997})}\BibitemShut {NoStop}%
\bibitem [{\citenamefont {Balkundi}\ and\ \citenamefont
  {Harrison}(2006)}]{balkundi:harrison:2006}%
  \BibitemOpen
  \bibfield  {author} {\bibinfo {author} {\bibfnamefont {P.}~\bibnamefont
  {Balkundi}}\ and\ \bibinfo {author} {\bibfnamefont {D.~A.}\ \bibnamefont
  {Harrison}},\ }\href@noop {} {\bibfield  {journal} {\bibinfo  {journal}
  {Acad.\ of Manag.\ Journal}\ }\textbf {\bibinfo {volume} {49}},\ \bibinfo
  {pages} {49} (\bibinfo {year} {2006})}\BibitemShut {NoStop}%
\bibitem [{\citenamefont {Efron}\ and\ \citenamefont
  {Tibshirani}(1993)}]{efron:tibshirani:1993}%
  \BibitemOpen
  \bibfield  {author} {\bibinfo {author} {\bibfnamefont {B.}~\bibnamefont
  {Efron}}\ and\ \bibinfo {author} {\bibfnamefont {R.~J.}\ \bibnamefont
  {Tibshirani}},\ }\href@noop {} {\emph {\bibinfo {title} {An Introduction to
  the Bootstrap}}}\ (\bibinfo  {publisher} {Chapman \& Hall},\ \bibinfo
  {address} {New York, NY},\ \bibinfo {year} {1993})\BibitemShut {NoStop}%
\bibitem [{\citenamefont {Hanley}\ and\ \citenamefont
  {McNeil}(1982)}]{hanley:mcneil:1982}%
  \BibitemOpen
  \bibfield  {author} {\bibinfo {author} {\bibfnamefont {J.~A.}\ \bibnamefont
  {Hanley}}\ and\ \bibinfo {author} {\bibfnamefont {B.~J.}\ \bibnamefont
  {McNeil}},\ }\href@noop {} {\bibfield  {journal} {\bibinfo  {journal}
  {Radiology}\ }\textbf {\bibinfo {volume} {143}},\ \bibinfo {pages} {29}
  (\bibinfo {year} {1982})}\BibitemShut {NoStop}%
\end{thebibliography}%

\end{document}



\title{Environmental structure and competitive scoring advantages in team competitions \\
\textit{Supporting Information}}


\author{Sears Merritt}
\email[]{sears.merritt@colorado.edu}
\affiliation{Department of Computer Science, University of Colorado, Boulder, CO 80309}

\author{Aaron Clauset}
\email[]{aaron.clauset@colorado.edu}
\affiliation{Department of Computer Science, University of Colorado, Boulder, CO 80309}
\affiliation{BioFrontiers Institute, University of Colorado, Boulder, CO 80303}
\affiliation{Santa Fe Institute, 1399 Hyde Park Rd., Santa Fe, NM 87501}


\begin{abstract}
\end{abstract}

\pacs{}

\maketitle


\renewcommand{\thefigure}{S\arabic{figure}}
\setcounter{figure}{0}
\renewcommand{\thetable}{S\arabic{table}}
\setcounter{table}{0}
\renewcommand{\thepseudocode}{S\arabic{table}}
\setcounter{pseudocode}{0}

\section{Detailed Description of Data}
\label{appendix:g}
\textit{Halo:\!\! Reach} is a popular online game played by nearly 20 million individuals, and was the 3rd most popular US video game of 2010~\cite{esa:2011}. It was publicly released by Bungie Inc., a former subdivision of Microsoft Game Studios, on 14 September 2010, and since then, players have generated more than 1 billion competitions. \textit{Reach} is an example of the kind of virtual combat simulation known as a ``first-person shooter'' or FPS. Within the \textit{Reach} system, players choose from among roughly seven primary game types and numerous subtypes, which are played on more than 33 terrain maps with 74 weapons (the precise number of maps and weapons has varied over time, as the publisher has periodically revised the online content through downloadable updates).

Instances of the game can be played alone, with or against other players via the Xbox Live online system. Participation in this system requires an account, which is distinguished by unique and publicly known ``gamertag'' or online pseudonym, chosen by the player. In the \textit{Reach} system, both individual game and player summaries were made publicly available through the Halo Reach Stats API. Through this digital interface, we collected detailed data on the first 53 million competition instances (roughly 1TB of data).

Within our sample, there are three basic game types: \emph{campaign games}, a sequence of story-driven, player-versus-environment (PvE) maps that many players complete first; \emph{firefight games} (also PvE), in which a team of human-controlled players battle successive waves of computer-controlled enemies; and \emph{competitive games}, a player-versus-player (PvP) game type, in which teams of the equal size (2, 4, 6 or 8 players) compete to either be the first to reach some fixed number of points or have the largest score after a fixed length of time. (The precise number of players per team, number of points required to win and length of a game depends on the game subtype.) Here, we focus on the most common type of competitive game, with teams of 4 players, a time limit of 600 seconds and a score limit of 50 points.

Among other information, each competition instance game file includes the sequence of scoring events at the per-second resolution and a list of players by team. Scoring events are annotated with the gamertag of the player generating the event, the number of points scored and the player giving up the points (if applicable).

Unlike professional sports, team composition and player resources in {\em Reach} competitions are not persistent across instances. The only attribute that persists is individual player skill, and thus each new instance is a kind of a ``blank slate.'' To join a new instance, individual players or small groups (often friends~\cite{mason:clauset:2012}) first enter a general pool of available competitors. A Bayesian ``matchmaking'' algorithm, which seeks to build teams of equal skill~\cite{herbrich:minka:graepel:2007}, then fills teams in the new instance by drawing from this pool. This process substantially randomizes the pairing of individuals within teams and the pairing of teams across instances. Because of the matchmaking algorithm and the large size of the pools, a pair of non-friend players are highly unlikely to be paired again in a new instance; friends may elect to be matched as a unit by forming a ``party,'' a special grouping that the matchmaking algorithm recognizes.

The non-persistence and the randomization are features absent from most studies of team performance or competition~\cite{ruef:aldrich:carter:2003,baldwin:etal:1997,balkundi:harrison:2006}, and serve to mitigate the confounding effects of persistent teams and resources present in most competitive systems, e.g., professional sports. For our purposes, these features make \textit{Reach} competitions a unique source of data for studying behavioral dynamics within competitions and how structural factors shape this behavior.

In competitive games, players move their avatars through the game map simultaneously, in real-time, navigating complex terrain, acquiring avatar modifications and encountering opponents. Teammates may interact through a private voice channel, or through visual signals. Points are scored by dealing sufficient damage to eliminate an opposing avatar and for each such success, a team gains a single point. Eliminated players must then wait several seconds before their avatar is placed back into the game at one of several specified ``spawn'' locations, equipped with ``default'' avatar resources that depend on the competition type being played.

For our analysis, we exclude all PvE games and all PvP games containing corrupt scoring event data. (Our analysis suggests no specific pattern to the corruption.) In our primary analyses, we further restricted our sample to PvP competitions (i) between two teams of 4 players and (ii) where no player exited the game early. This latter criterion was relaxed to calculate the relationship between dropouts and $\beta$ (see Section~\ref{appendix:i}).

\section{Generative Model for Scoring Event Timing and Balance}
\label{appendix:a}
The timing and balance (which team receives the point) of scoring events within a competition are modeled by a conditionally independent Markov process, where an incremental change to a team's score $s_{r}$ is given by 
\begin{align}
\Pr(\Delta s_{r}(t) > 0) = \Pr(\Delta s_{r}>0\,|\,\theta,\textrm{event}) \Pr(\textrm{event at }t\,|\,\theta\,) \nonumber \enspace ,
\end{align}
where $\theta$ parameterizes the impact of non-ideal competitive features. That is, the probability that team $r$'s score increases at some time $t$ is the probability that a scoring event occurred at time $t$ and that the resulting point was awarded to $r$. Furthermore, team labels $r$ and $b$ are arbitrary, and we choose $r$ as our reference team below.

The generation of scoring events is given by a non-stationary Poisson process, in which the probability that a scoring event occurs at time $t$ varies linearly with time:
\begin{align}
\Pr(\textrm{event at }t\,|\,\lambda_{0},\alpha\,) & =  \lambda_{0}+\alpha\,t \enspace ,
\end{align}
where $\lambda_{0}$ is the event background rate and $\alpha$ is the acceleration. When $\alpha=0$, we recover the stationary Poisson process expected for ideal competitions.

In a real competition, we observe $n\leq T$ scoring events, for a competition lasting $T$ units of time. Let $\{ t_{i} \}$ denote the observed times of these events, and $\{u_{j}\}$ the times at which no event was observed. The model parameters $\lambda_{0}$ and $\alpha$ are then jointly estimated by directly maximizing the generative model's log-likelihood function:
\begin{align}
\ln \mathcal{L} & = 
\sum_{i=1}^{n} \ln(\lambda_0+\alpha\, t_i) + \sum_{j=1}^{T-n} \ln(1-\lambda_0-\alpha\, u_j) \enspace . \label{eq:scoring:like}
\end{align}

\begin{figure*}[t!]
\centering
\begin{tabular}{cc}
\includegraphics[scale=0.45]{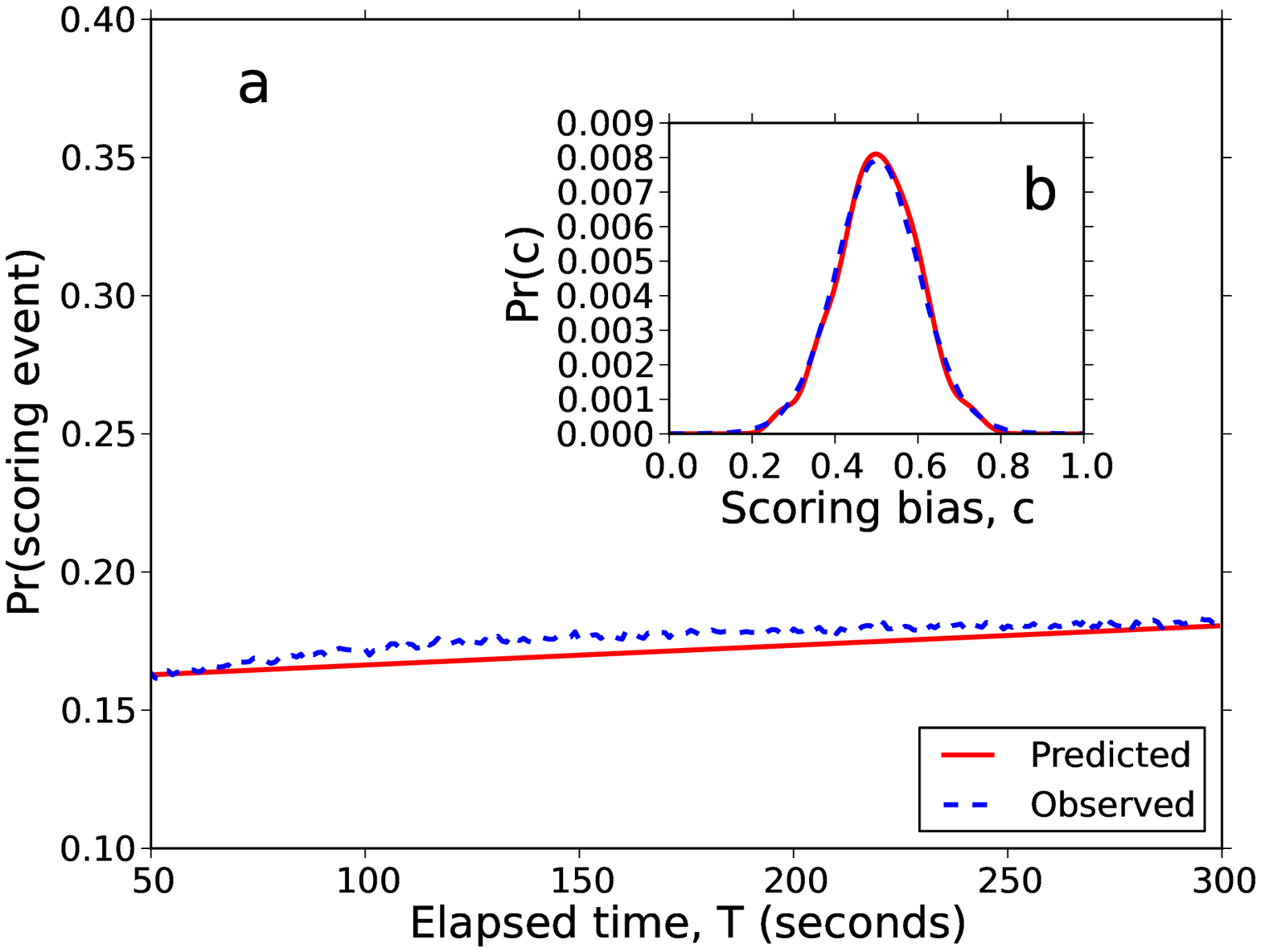} &
\includegraphics[scale=0.45]{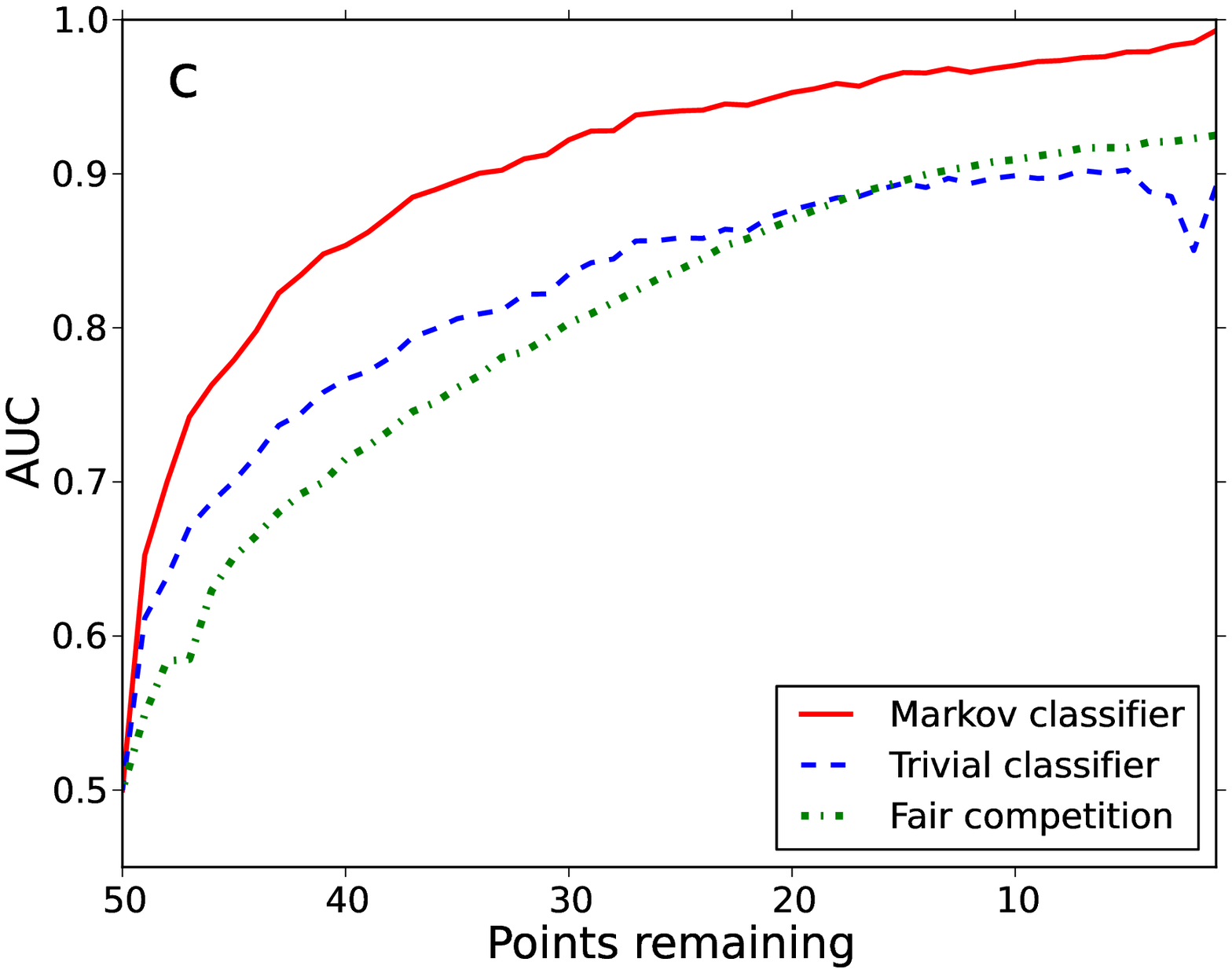}
\end{tabular}
\caption{(A) Global empirical and predicted scoring rates for competitions in \textit{Halo:\!\! Reach}, over the window $[50,300]$ seconds. (A, inset) Global empirical and predicted distribution of competitive advantages (smoothed via a Gaussian kernel). (B) For all competitions, winner predictability (AUC) as a function of team $r$'s points remaining, for three classifiers (see text).}
\label{fig:param:fits:auc}
\end{figure*}

To limit the biasing effect of the highly non-stationary behavior found in the early- and end-phases of competitions (see main text), we restrict our estimation to events occurring in the middle phase, specifically $50 \leq t \leq 300$. This heuristic provides robust conclusions: the estimated timing parameters are very close to those found using smaller middle-phase windows, and the global average trend within this window is roughly linear (Fig.~\ref{fig:param:fits:auc}A).

For two teams $r$ and $b$, the outcome of a scoring event (which team receives the point) is given by a biased Bernoulli process, in which the probability that an event increases the score of team $i$ is
\begin{displaymath}
\Pr(s_{i}\textrm{ increases}\,|\,\theta\,) = \left\{ 
\begin{array}{ll}
c & i=r \\
1-c & i=b \enspace , \\
\end{array}
\right.
\end{displaymath}
where $c \in [0,1]$ represents the competitive advantage (outcome bias) of the $r$ team. In our model system, 99.99\% of scoring events yield a single point. Although we do not consider the possibility here, in general, the number of points produced by an event could be drawn from some distribution. Thus, the probability that the competition ends with final scores $S_{r}$ and $S_{b}$ is
\begin{align}
\Pr(S_{r}, S_{b} \,|\, c ) & = c^{S_{r}}(1-c)^{S_{b}} \enspace , \label{eq:outcome:like}
\end{align}
where $c$ denotes the competitive advantage (scoring bias) of team $r$ over team $b$.

Because team composition varies across competition instances, the competitive advantage of $r$ is modeled as a random variable, drawn from some distribution $\Pr(c)$. The natural choice of the form of this distribution is 
a symmetric Beta distribution with parameter $\beta$, the conjugate prior for the Bernoulli scheme. 
(We note that the prior distribution must be symmetric about $c=1/2$ because team labels are arbitrary.)
This distributional assumption agrees well with the global empirical distribution of biases $c$ (Fig.~\ref{fig:param:fits:auc}A inset). 

The posterior probability of observing final scores $\{ S_{r}, S_{b}\}_{k}$ in a competition instance $k$ is given by their Bernoulli likelihood, weighted by the probability of $c$  (Eq.~\eqref{eq:outcome:like}). Given $N$ such instances, the total posterior probability of the observed final scores is
\begin{align}
\Pr(\beta\,|\,\{S_{r},S_{b}\}) & = \int_{0}^{1} \left( \prod_{k=1}^{N}\Pr(\{S_{r},S_{b}\}_{k}\,|\, c)  \Pr(c \,|\, \beta) \right) \, \textrm{d}c \nonumber \\
 & = \prod_{k=1}^{N} \left( \int_{0}^{1} \frac{c^{S_{r_k}+\alpha-1} (1-c)^{S_{b_k}+\beta-1}}{\textrm{B}(\beta,\beta)} \, \textrm{d}c\right)  \nonumber \\
& = \prod_{k=1}^{N} \frac{\textrm{B}(S_{r_k}+\beta,S_{b_k}+\beta)}{\textrm{B}(\beta,\beta) } \enspace , \label{eq:outcome:posterior}
\end{align}
where $\textrm{B}(a,b)$ is the Beta function.

We estimate the competition balance parameter by numerically maximizing the logarithm of Eq.~\eqref{eq:outcome:posterior} with respect  to $\beta$,
\begin{align}
\ln \mathcal{L} & = 
\sum_{k=1}^N \ln[\textrm B( S_{r_k} + \beta, S_{b_k} + \beta)] - \ln[\textrm B(\beta, \beta)] \label{eq:skill:ll} \enspace .
\end{align}
The resulting maximum likelihood estimate $\hat{\beta}$ provides a direct measurement of the overall balance within a set of competition instances: when $\beta\to\infty$, we recover the fair coin $c=1/2$ expected for ideal competitions.

For a set of competition instances, numerically maximizing Eq.~\eqref{eq:scoring:like} with respect to $\lambda_{0}$ and $\alpha$, and Eq.~\eqref{eq:skill:ll} with respect to $\beta$, produces maximum likelihood parameter estimates ${\hat \lambda_0}$, ${\hat \alpha}$, and ${\hat \beta}$. Uncertainty in these estimates is then calculated as the standard deviation of the bootstrap distribution~\cite{efron:tibshirani:1993}, where we resample compelte competition instances with replacement. Table~\ref{tab:global:param:est} gives the global parameters estimates and uncertainties, when applied to the full set of \textit{Halo:\!\! Reach} competitions.



\begin{table}[t]
\begin{tabular}{clrl}
\multicolumn{2}{l}{parameter} &  \multicolumn{2}{c}{estimate, global} \\
\hline
$\beta$ & balance &  $29.50$ & $\pm~0.21$\\
$\lambda_0$ & base rate &  $0.1620$ & $\pm~0.0001$\\
$\alpha$ & acceleration &  ~~$7.00\times10^{-5}$ & $\pm~0.05\times10^{-5}$ \\
\hline
\end{tabular}
\caption{Estimated global scoring tempo and balance parameters, with bootstrap uncertainty estimate.}
\label{tab:global:param:est}
\end{table}%

\section{Predicting Competition Outcomes}
\label{appendix:c}
For a set of competitions, the predictability of an instance's ultimate winner, after observing only part of the game, provides a second, non-parametric measure non-ideal dynamics. We model scoring as a Markov chain that terminates when a team reaches a score of 50. (In our data, 99\% of  competitive instances terminate according to this criteria; the remainder from the time limit.)

Suppose an instance has evolved so that teams $r$ and $b$ currently hold scores $s_r$ and $s_b$. The probability that team $r$ wins the competition is then
\begin{align}
\Pr(r \textrm{ wins} \,|\, s_r, s_b) = & \Pr(r \textrm{ wins} \,|\, s_r+1, s_b)\cdot {\hat c} \,+ \nonumber \\
 &  \Pr(r \textrm{ wins} \,|\, s_r, s_b+1) \cdot (1-{\hat c}) \enspace ,
\label{eq:predict}
\end{align}
where ${\hat c} = s_r/ (s_r+s_b)$ is the current maximum likelihood estimate of $r$'s scoring bias within this instance, and the two probability terms capture the probability that $r$ wins if $r$ (or $b$) wins the next point. (Because a team's score is cumulative, each state in the Markov chain has only two transitions.) Eq.~\eqref{eq:predict} is then solved recursively by computing ${\hat c}$ for the current state and working backwards to the instances's current state from the winning states where $s_r=50$ and $s_b<50$.

We convert this Markov chain into a classifier by predicting that team $r$ wins if $\Pr(r \textrm{ wins} \,|\, s_r, s_b) > 0.5$. The probability of correctly choosing the winning team in this case is equivalent to computing the AUC statistic over a set of instances. (AUC is defined as the area under the receiver-operating characteristic (ROC) curve~\cite{hanley:mcneil:1982}, and is mathematically equivalent to the Mann-Whitney $U$ test for distinguishing two classes of items.) 

Measuring the AUC as a function of the points remaining provides full information about the way the competition's predictability evolves over time. We convert this information into a point measure by computing, with 40 points remaining for $r$, the AUC for the Markov classifier, which we then divided by the corresponding AUC for an ``ideal'' classifier (with fixed $c=1/2$). This provides a direct measure of how much more predictable a real competition's outcome is relative to the ideal model described in the main text.

Using the full data set, Figure~\ref{fig:param:fits:auc}B shows the full AUC-over-time curves, for the Markov classifier, the ideal classifier ($c=1/2$), and for a trivial classifier in which at each moment we predict as the winner the team currently in the lead. Our Markov classifier outperforms the trivial classifier because it captures information about the size of the lead, i.e., it includes information about the bias $c$ in the Bernoulli scoring process, and outperforms the ideal classifier because the competitions' dynamics are non-ideal.


\begin{figure}[t!]
\begin{center}
\includegraphics[scale=0.40]{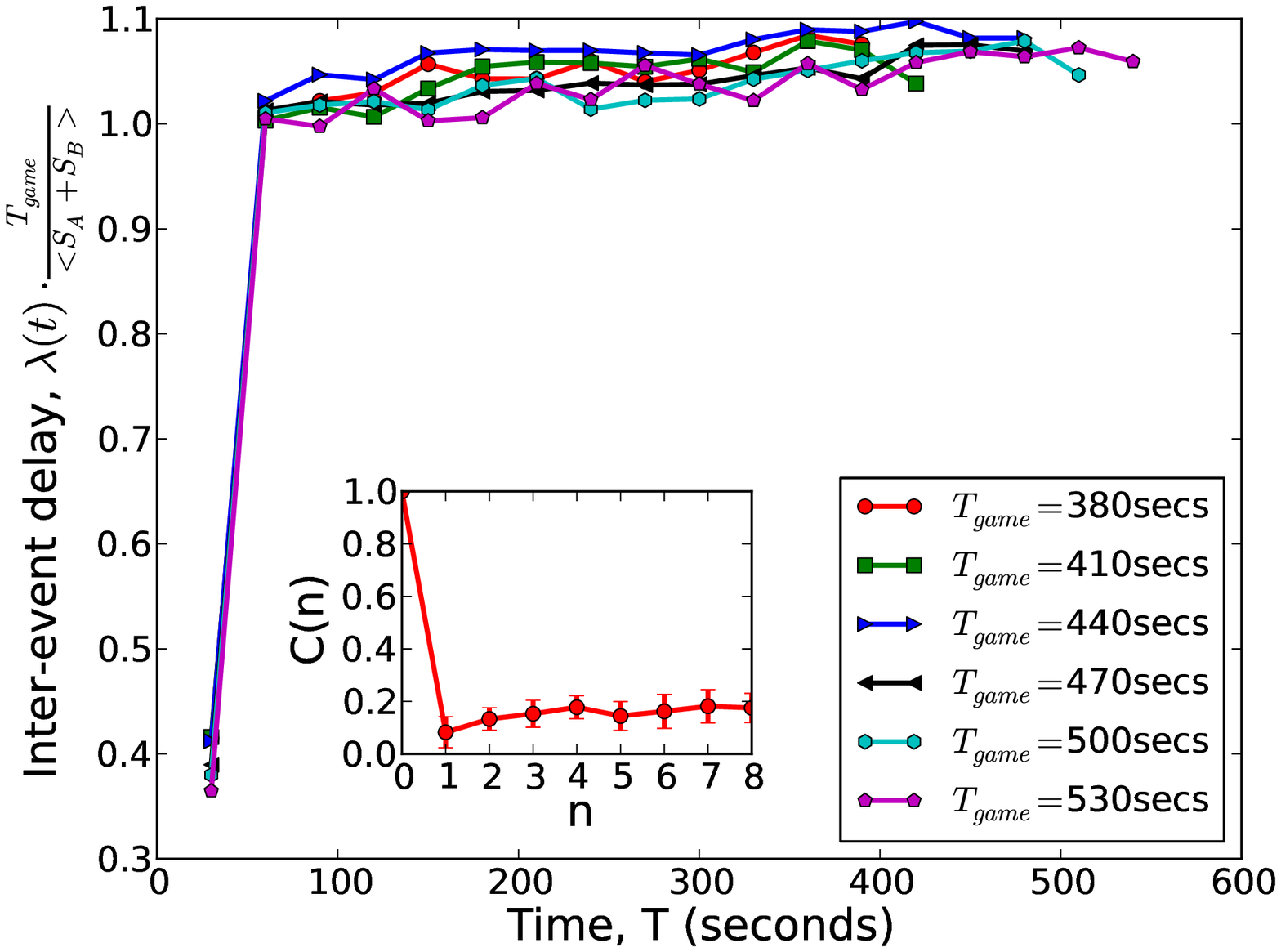} 
\caption{Average normalized inter-arrival time between scoring events, computed in 30 second intervals, for cohorts of competitions lasting a specific amount of time. (inset) Auto-correlation function $C(n)$ for inter-event times.}
\end{center}
\label{fig:lambda:over:time}
\end{figure}

\begin{figure*}[t!]
\centering
\begin{tabular}{cc}
\includegraphics[scale=0.38]{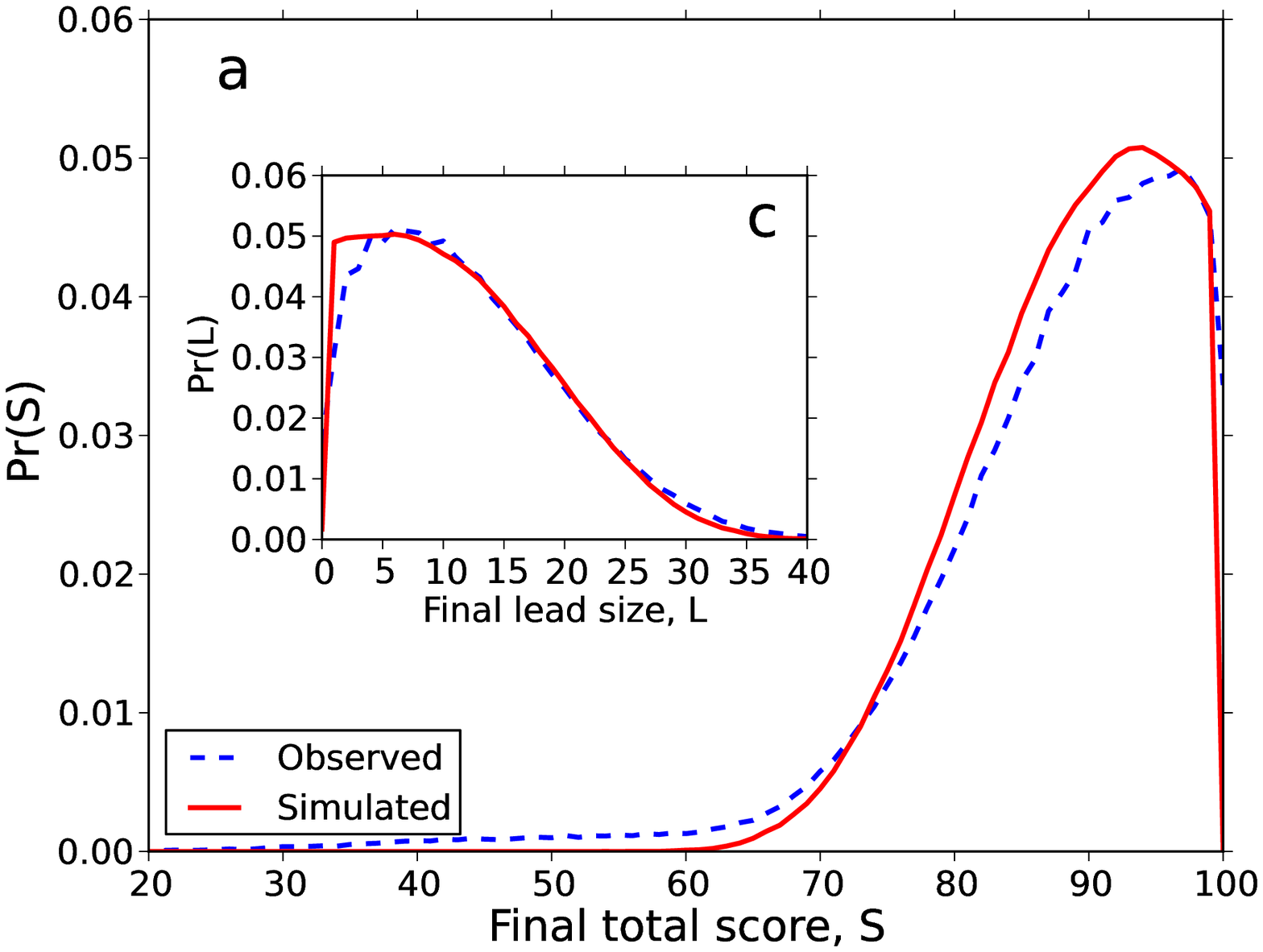} & 
\includegraphics[scale=0.38]{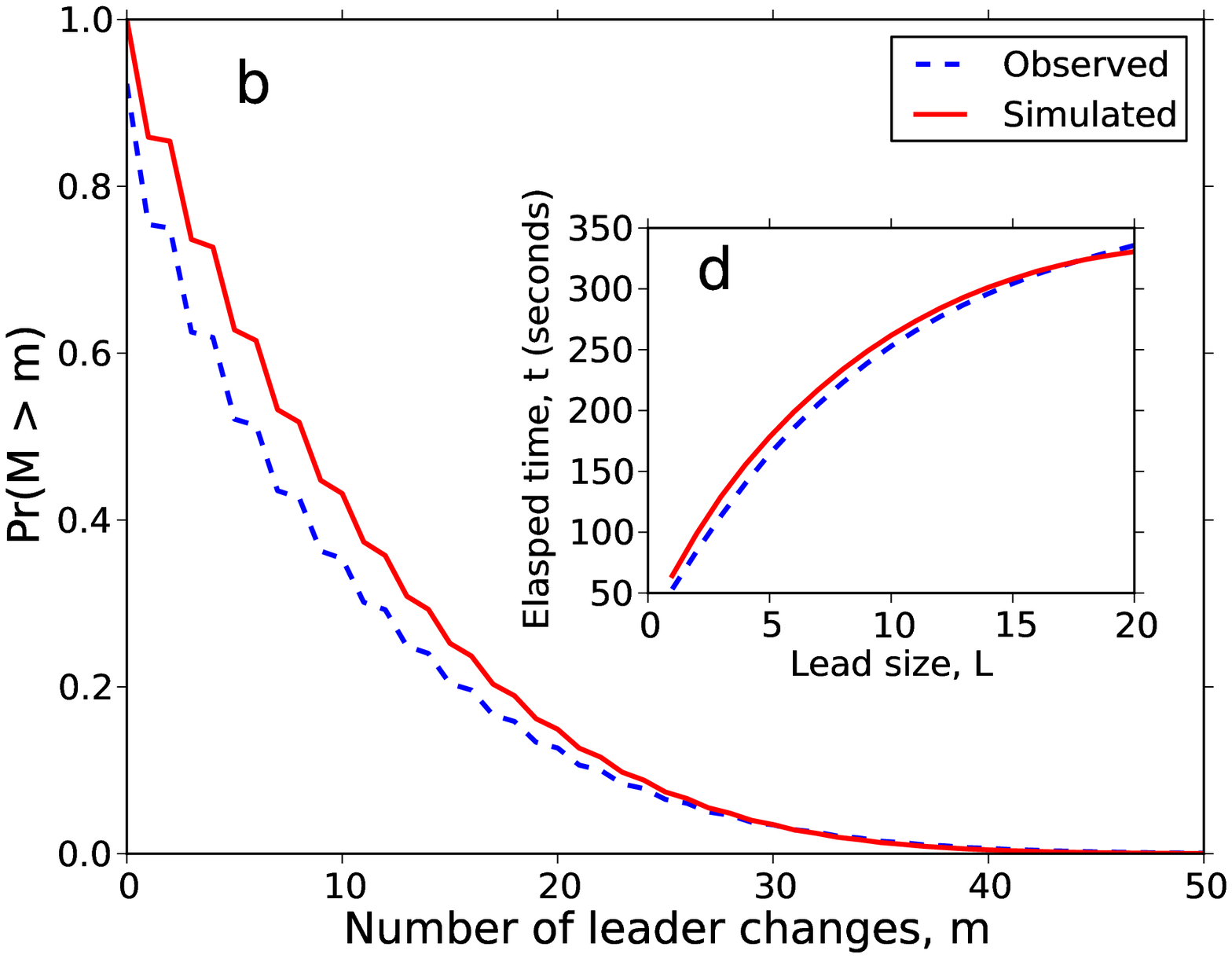}
\end{tabular}
\caption{Comparison of empirical (dashed blue) and simulated (parametric model, red) data for the (A) distribution of final total scores $S=S_{r}+S_{b}$, (B) distribution of the number of times the identity of the leading team changes $m$, (C) distribution of final lead sizes $L=|S_{r}-S_{b}|$, and (D) time $t$ elapsed as leader given a lead size of $L$. The close agreement between data and simulation suggests that our generative model efficiently captures these competitions' dynamics.}
\label{fig:final:scores:leads}
\end{figure*}

\section{Test of the Markov Assumption}
\label{appendix:d}
We now test the accuracy of our Markov assumption in modeling the scoring dynamics of these competitions. If the arrival times of scoring events roughly follow a memoryless Poisson process, there will be little correlation between the sizes of subsequent delays. The correlation function $C(n)$ provides a direct measure of the accuracy of the Markov assumption, and is calculated as
\begin{align}
C(n) & = \frac{\langle T_{i} T_{i+n}\rangle - \langle T_{i} \rangle^{2}}{\langle T_{i}^{2}\rangle - \langle T_{i}\rangle^{2}} \enspace ,
\label{eq:correlation}
\end{align}
where $T_{i}$ is the inter-event delay after event $i$, $n$ is a shift size relative to $i$, and $\langle . \rangle$ indicates an average over $i$. A memoryless process matching the Markov assumption in our Bernoulli process will produce $C(n)\approx 0$ for $n>0$; deviations indicate correlations (or anti-correlations) at the corresponding time scale. 

First, a simple rescaling of the observed inter-event delays over the course of competitions of different lengths produces a data collapse (Fig.~S2), illustrating relatively little memory in the system. Second, $C(n)$ for our entire sample of competitions (Fig.~S2, inset) shows little correlation (memory) at any time scale. Thus, the Markov assumption seems largely justified.

\section{Model Goodness-of-Fit}
\label{appendix:f}

We now test the plausibility of our generative model, i.e., how well it matches the underlying data, by comparing simulated competitions against the empirical data along specific statistical measures. This simulation is parametric and uses the estimated parameters from our generative model to define the corresponding probability distributions in the simulator. A close match between the synthetic scoring dynamics and the empirical data along multiple statistical measure is evidence that our generative model accurately captures the basic features of these competitions.

The simulation framework is given in Algorithm~\ref{algorithm:sim}. The competition clock is started at $t=25$ seconds to account for the early-phase delay in the onset of scoring. The bias in the Bernoulli process is then chosen by drawing a value iid from the estimated Beta distribution with parameter $\hat{\beta}$. While neither of the termination criteria have been reached, delays between scoring events are drawn from the estimated linear non-stationary process with parameters $\hat{\lambda}_{0}$ and $\hat{\alpha}$. Finally, given that a scoring event occurs, with probability $c$, a single point is awarded to team $r$; otherwise, it is awarded to $b$.

\begin{pseudocode}{Competition simulation}{ }
t \GETS 25 \\
s_r \GETS s_b \gets 0 \\
c \GETS \textrm{chooseScoringBias()} \\
\WHILE{t < 600 \AND s_r < 50 \AND s_b < 50} \DO
	\BEGIN
	T \GETS \textrm{interEventDelay()} \\
	\IF t + T < 600
		\THEN 
		\BEGIN
		\Delta s \GETS \textrm{numPoints()} \\
		\textrm{updateScores($s_{a}$,$s_{b}$, $\Delta$ s,c)} \\
		t \GETS t + T  \\
		\END
	\ELSE \textrm{break}
	\END
\label{algorithm:sim}
\end{pseudocode}

The goodness-of-fit of the model is measured by comparing the simulated and empirical distributions of (i) the final score $S$, (ii) the final lead size $L$ (at termination), (iii) the number of leader changes $m$, and (iv) the amount of time $t$ the leading team stays in the lead given a lead of size $L$. Notably, each of these four quantities is distinct (although related) to the aspects of the data used to estimate the parametric model's structure, and thus they make reasonable checks on the accuracy of the model. Figures~\ref{fig:final:scores:leads}A-D show the results of these tests, using 1 million simulated competitions, illustrating very good agreement on all dimensions between simulation and data. Thus, the basic structure of our generative model seems largely justified.

\section{Additional Results for How Structure Shapes Dynamics}
\label{appendix:e}
\begin{figure*}[t!]
\centering
\begin{tabular}{cccc}
\includegraphics[scale=0.22]{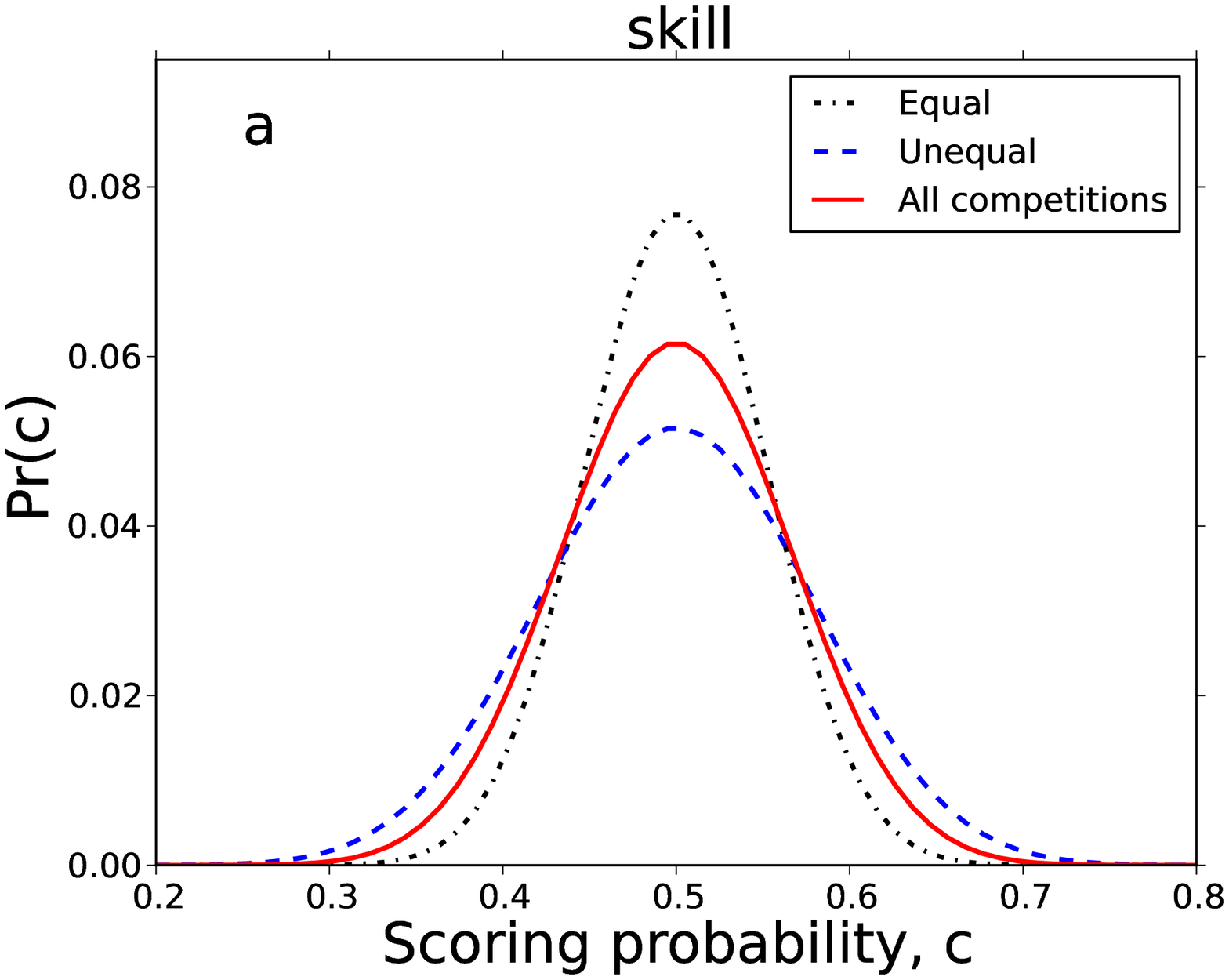} &
\includegraphics[scale=0.22]{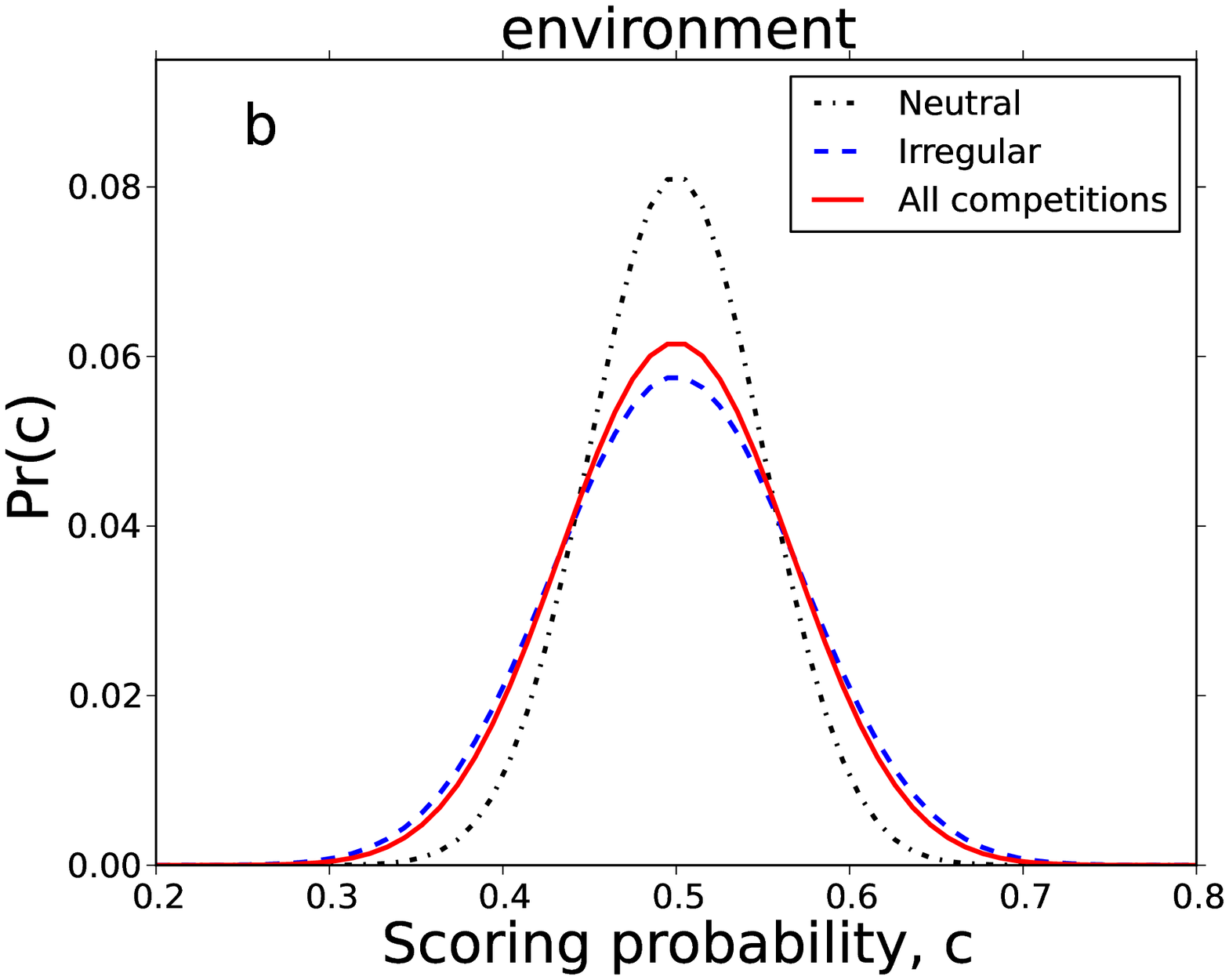} &
\includegraphics[scale=0.22]{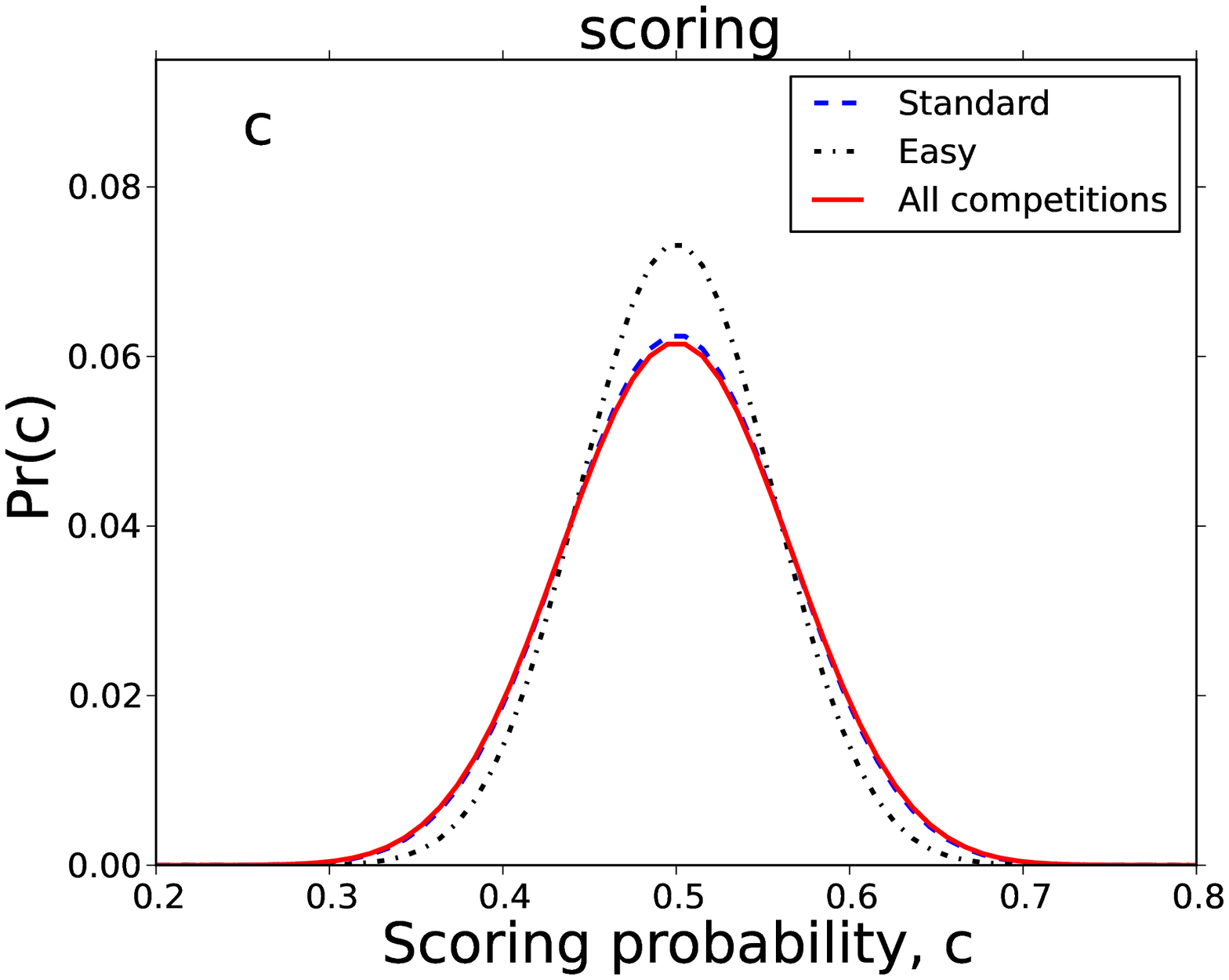} &
\includegraphics[scale=0.22]{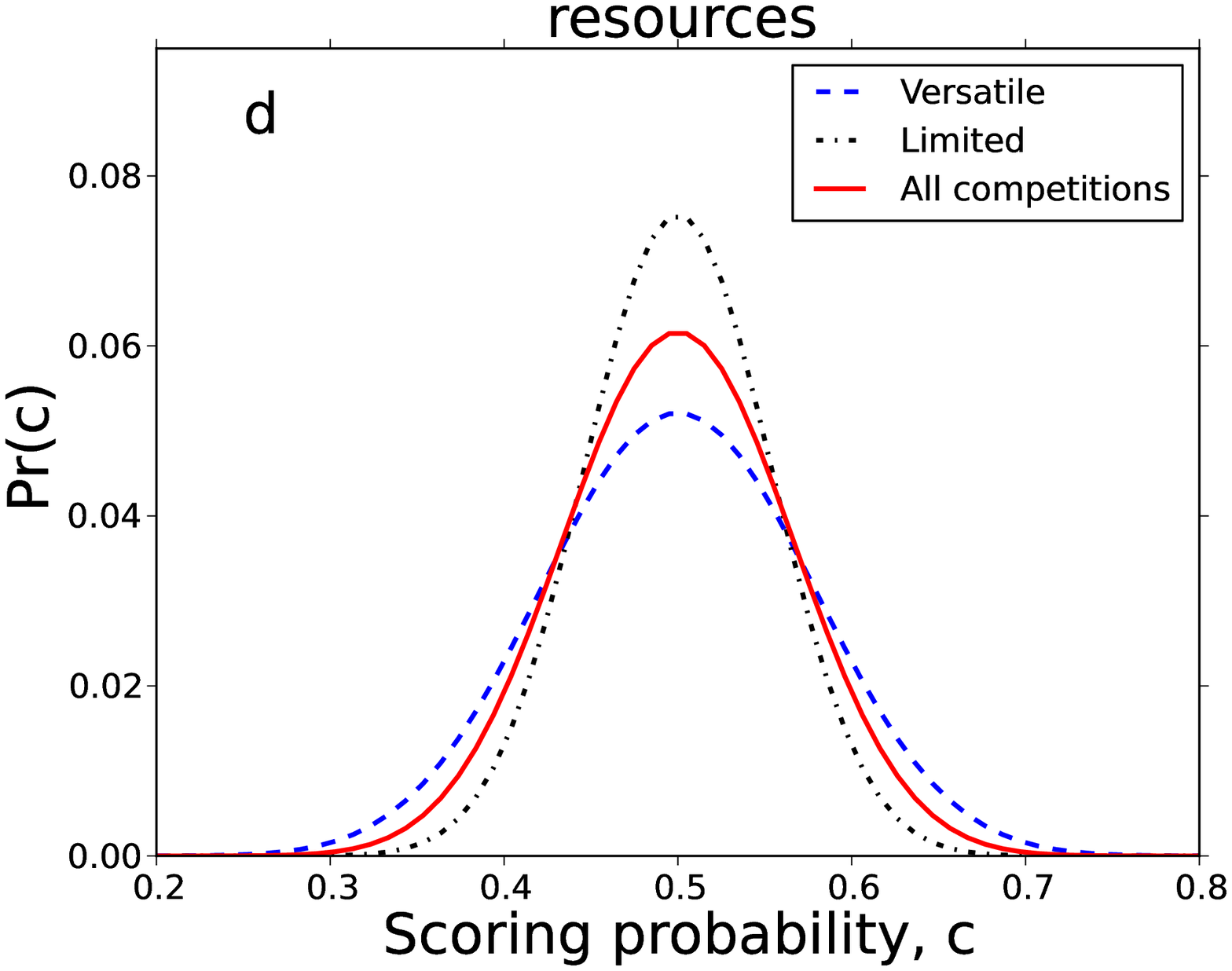} \\
\includegraphics[scale=0.22]{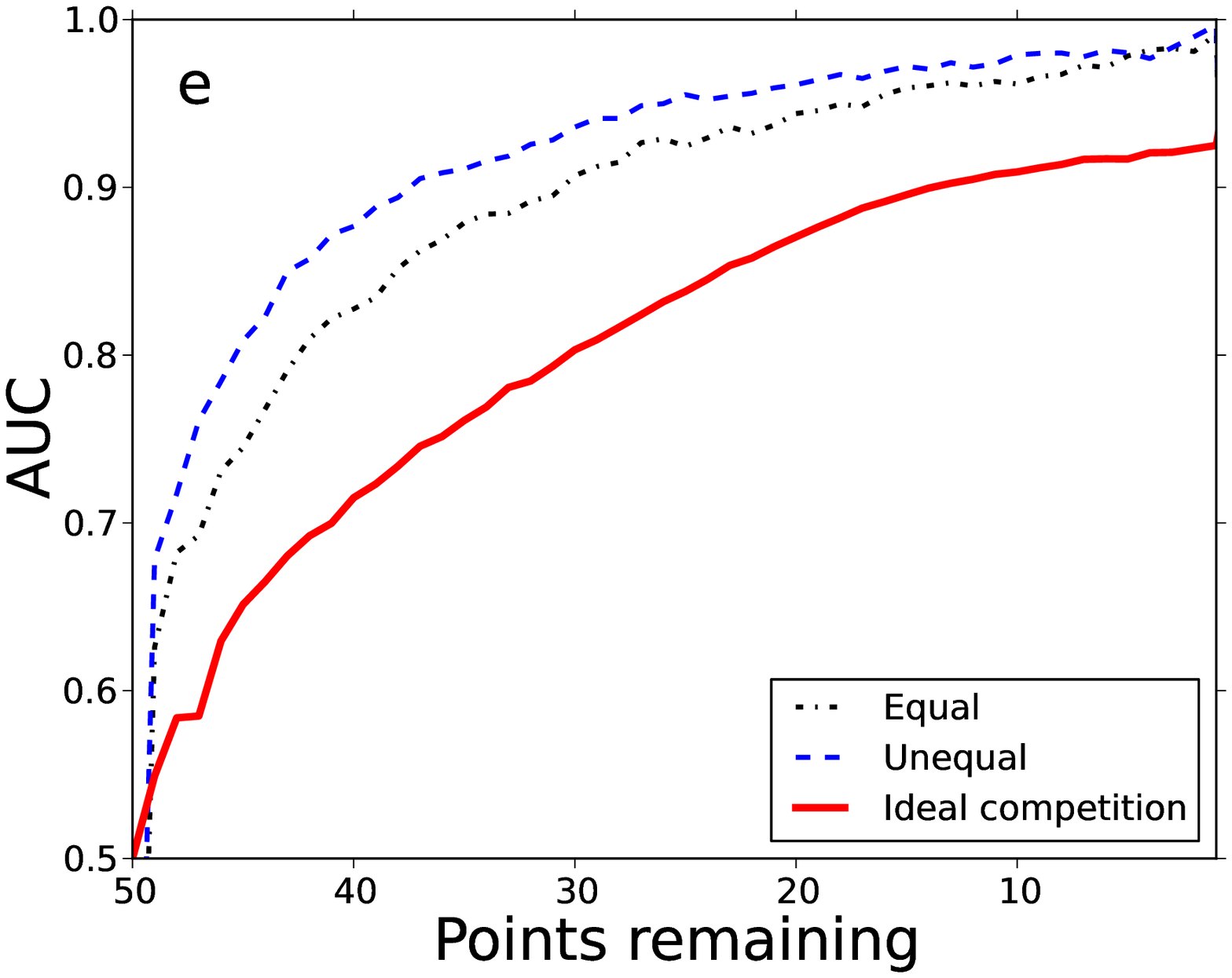} &
\includegraphics[scale=0.22]{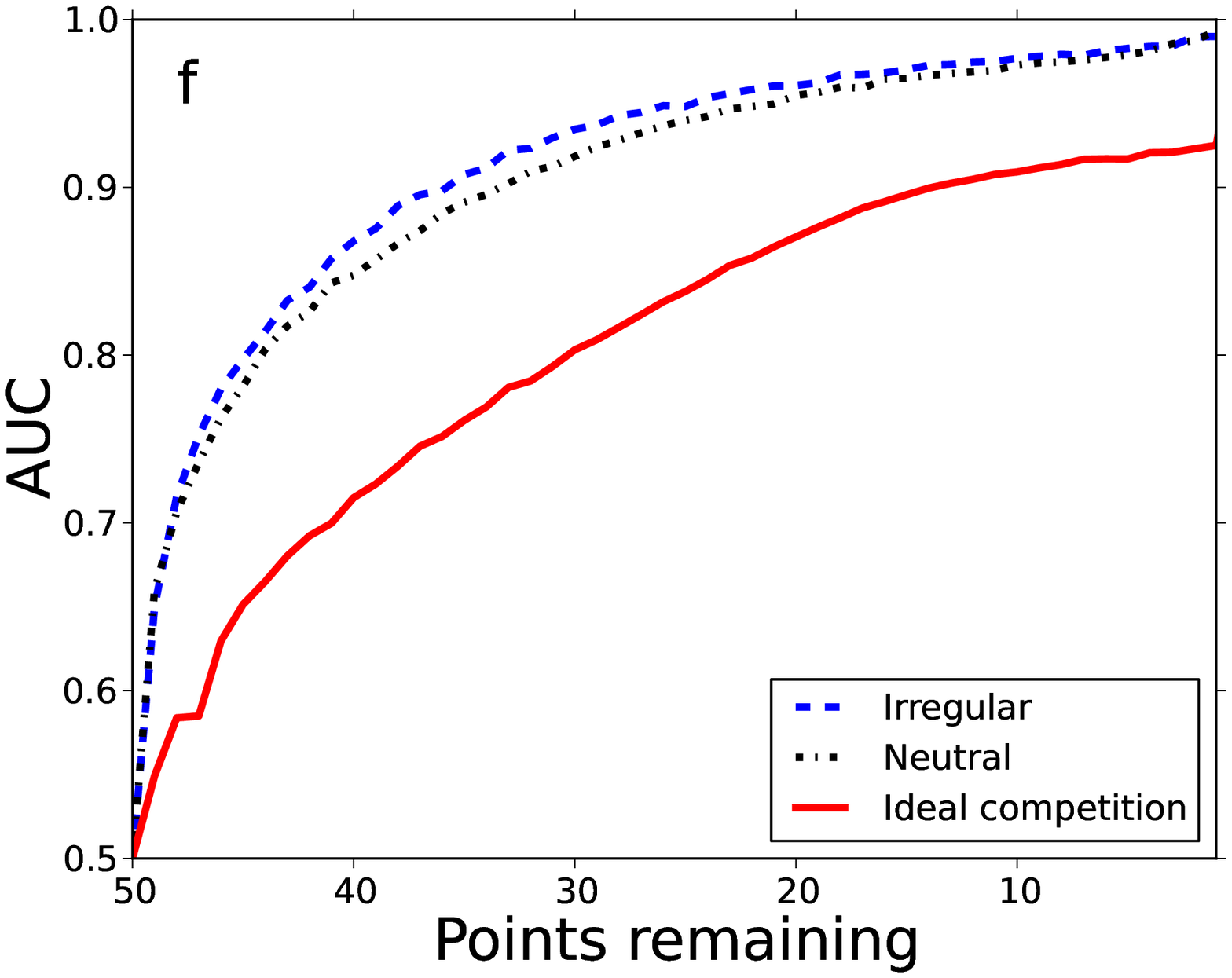} &
\includegraphics[scale=0.22]{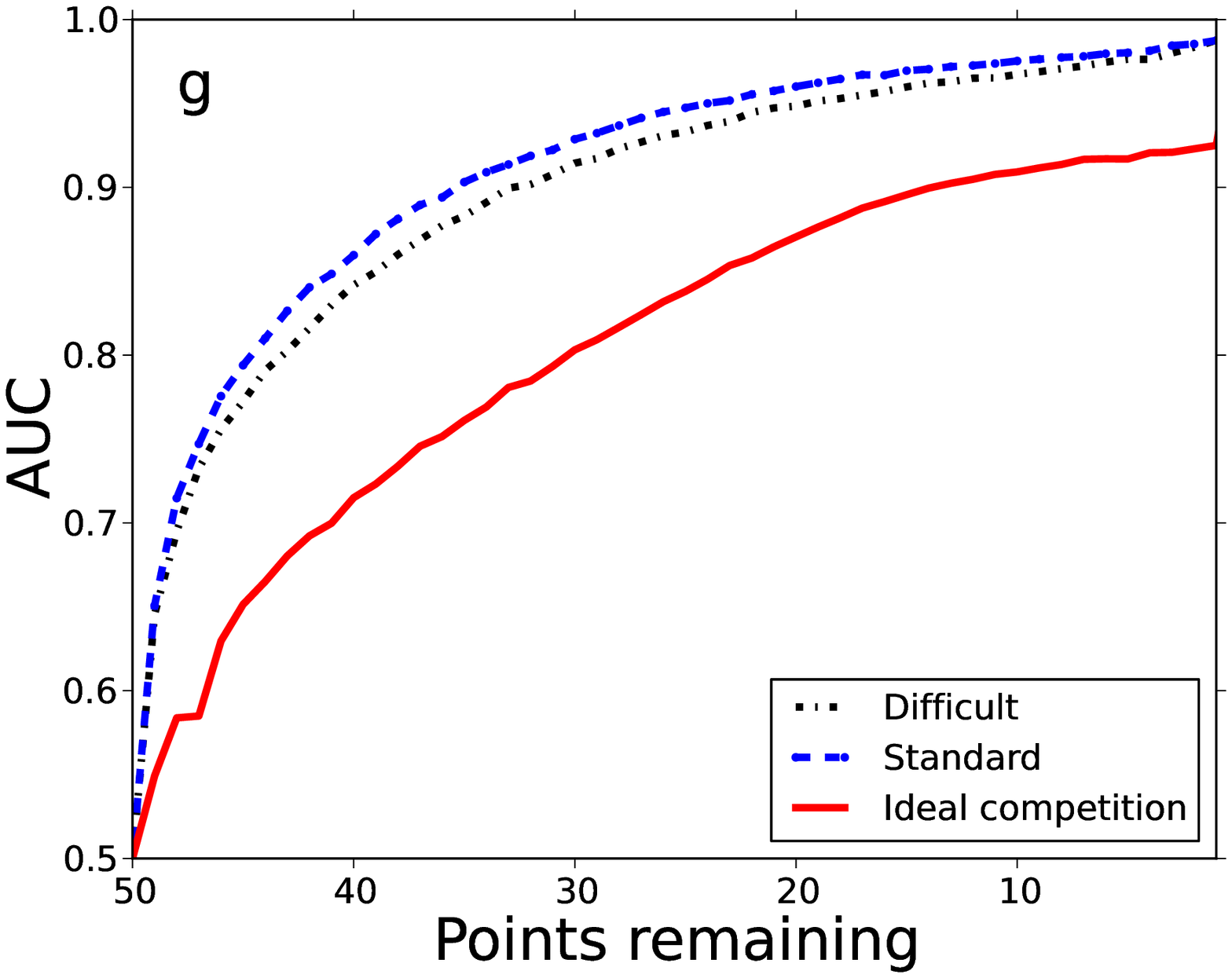} &
\includegraphics[scale=0.22]{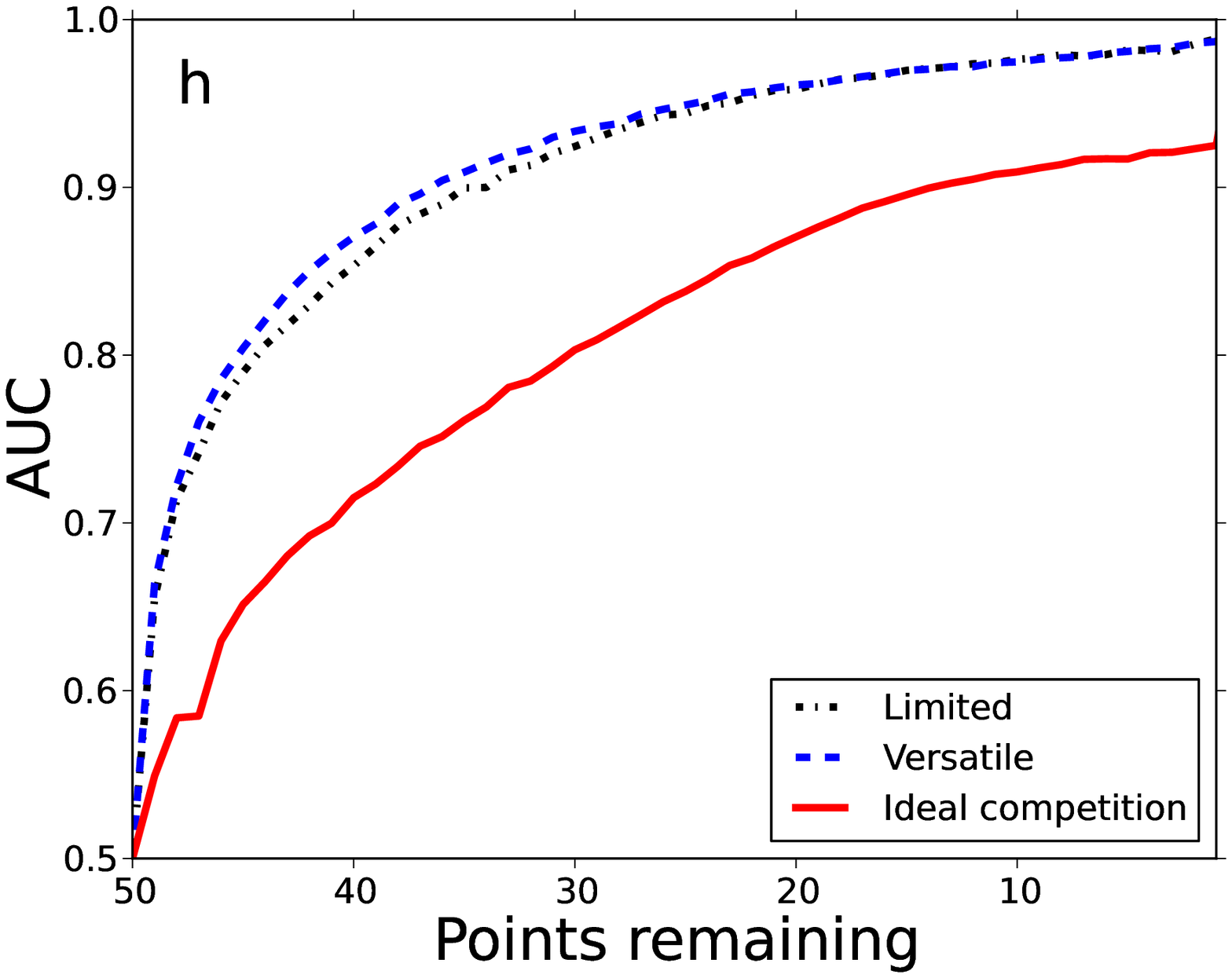}
\end{tabular}
\caption{For the four dimensions discussed in the main text, (A, B, C, D) estimated distribution of scoring biases $\Pr(c)$, and (E, F, G, H) the AUC as a function of points remaining in the competition.}
\label{fig:vignettes}
\end{figure*}

In the main text, we examined four pairs of competition types that each differed on one structural feature: team skill, environmental structure, policies, and resource quality. Figures~\ref{fig:vignettes}A-D show the estimated distributions of $\Pr(c)$ (parameterized by $\hat{\beta}$) for these four pairs. For each group of instances, the model parameter $\beta$ was estimated following Section~\ref{appendix:a} from the scoring events on the interval $t\in[30,300]$ seconds of the competition. These times were chosen to exclude biases due to early- and end-phase boundary effects.

Figures~\ref{fig:vignettes}E-H show the AUC as a function of points remaining for same competitions, estimated following Section~\ref{appendix:c}. In each figure, we show for comparison the AUC curve for an ideal competition ($c=1/2$). The large gap between the Markov classifier's AUC curve and the ideal curve demonstrates that these competitions are substantially more predictable than ideal competitions. This gap is largest early in the competition, where scores are still relatively far from the scoring limit. We also observe modest gaps between the AUC curves for members of each pair, illustrating that structural features do impact the predictability of competition outcomes.
\section{Additional details of multivariate regression analysis}
\label{appendix:h}
Here we describe additional details of our investigation of how resources, policy, environment, and skill features explain the variance in the values $\beta$, $\lambda_0$, $\alpha$, and $\rho$ observed in our data. To quantify the structure of a competition type $\vec{\eta}$, we defined 35 structural features that characterize the different combinations of environment, resources, policies, and teams. 
Table~\ref{tab:features} gives the full list of features, with descriptions, classified into four types: resources (R), environment (E), policies (P), and skill (S). Applied to our data yields 125 unique competition types (see Table~S10).

For all competition instances with a particular set of features, we estimated the coordinates $(\beta, \lambda_0, \alpha, \rho)$ following Sections~\ref{appendix:a} and~~\ref{appendix:c}.
Regression models were built on each coordinate independently, and robustness checks were conducted to verify these results (see below). Table~\ref{tab:lin:reg} lists the statistically significant ($p\le0.1$) features and corresponding coefficients for all four of our models. 

For competition balance $\beta$, we first used a linear model $\mathbf{\beta} = \mathbf{\theta}^T\mathbf{x}$, with a design matrix $\mathbf{x}$ composed of the previously defined 125 observations containing 35 features. Fitting this model via least squares produced $r^2=0.716$ ($p\ll0.001$, F-test), but with strongly skewed residuals. We then fitted the model $\log\mathbf{\beta} = \mathbf{\theta}^T\mathbf{x}$ to the data, which produced $r^2=0.933$ ($p\ll0.001$, F-test), a marked improvement, and more symmetric residuals.
Examining the coefficients, we find that evenly matched teams using medium-to-long-range weapons, competing on large environments without strategic or defensible positions produce more balanced scoring outcomes (larger $\beta$).

For the base scoring rate $\lambda_0$, a simple linear model yields $r^2=0.955$ ($p\ll0.001$, F-test), indicating that structural features explain almost all the observed variance. 
The estimated coefficients show that environmental structure features play a dominant role in setting $\lambda_0$. In particular, environments that are small, open, and circular correlate best with base scoring rate. In addition to the environment's spatial organization, evenly matched teams also correlate with higher scoring rates. Teams with more experience are likely to be familiar with all terrain options and methods for its exploitation. Environments that are small do not require competitors to spend much time seeking out scoring opportunities (other avatars). Lastly, environments that are open do not provide places to avoid encounters, thus increasing the tempo of competition.


For the acceleration $\alpha$ in the competition tempo, a linear model produces an $r^2=0.652$ ($p\ll0.001$, F-test). 
We find that few of our features correlate with $\alpha$, with the exception of long-range weapons and equally-skilled teams, which correlate with smaller $\alpha$ (more ideal competitions). This suggests that in competitions where players are experienced, there is less to learn and thus $\alpha$ is low. This agrees well with the results from $\lambda_0$, where more experience leads to a higher base scoring rate.

For the winner predictability $\rho$, a linear model produces an $r^2=0.885$ ($p\ll0.001$, F-test). 
Notably, features related to neutral environments and equally-skilled teams correlated with less predictable (more ideal) outcomes. As expected from the correlation between $\beta$ and $\rho$ (Table~\ref{tab:param:correlations}), features that correlated with greater $\beta$ typically also correlate with lower $\rho$. 

Finally, we expected changes in policy to have an impact on scoring balance and tempo of events. However, we find that policy type features do not by themselves play a role in controlling these dynamics, once we control for other variables like skill, environmental structure and resources. Specifically, we find that the policy feature coefficients are insignificant in all of our models ($p>0.1$) and thus we excluded from the results of our best-subset selection.

\subsection*{Tests of model robustness}
To test the robustness of our results against spurious correlation, due to the high-dimensionality of our data, we conducted three additional analyses.

First, we consider colinearity among the dependent variables. Table~\ref{tab:param:correlations} lists the pairwise coefficients of variation $r^{2}$, showing a high degree of correlation between $\rho$ and $\log \beta$, modest correlation between $\log \beta$ and $\lambda_{0}$, but little else. To test whether these correlations impact our results, we conducted a MANOVA on a multiple multivariate regression model (Table~\ref{tab:manova}). The results show that the same set of features reported in Table~\ref{tab:lin:reg} are significant, suggesting that our original results are robust.

Second, we perform a stepwise AIC feature selection procedure to choose the best subset of features under mild regularization. With the exception of $\alpha$, the results shown in Tables~\ref{tab:aic:a},~\ref{tab:aic:b}, and~\ref{tab:aic:c} indicate that the selected features and their weights presented in the original regression analysis are robust. The best-subset selection for $\alpha$ produces a larger list of significant features than in the original model, but a slightly lower $r^2$. The most significant negative feature, long range resources, is robust to this procedure while equally skilled teams and other resource features are not.

\begin{table}[t!]
\begin{center}
\begin{tabular}{cccccc}
 & $\log\beta$ & $\lambda_0$ & $\alpha$ & $\rho$ \\ \hline
$\log\beta$ & -- & 0.356 & 0.053 & 0.776 \\
$\lambda_0$ & 0.356 & -- & 0.003 & 0.398 \\ 
$\alpha$ & 0.053 & 0.003 & -- & -- \\
$\rho$ & 0.776 &  0.398 & -- & -- \\ \hline
\end{tabular}
\end{center}
\caption{Coefficients of variation $r^{2}$ for pairs of dependent variables. Cells containing no data are either irrelevant or statistically insignificant ($p > 0.1$).}
\label{tab:param:correlations}
\end{table}%

\begin{table}[t!]
\begin{center}
\begin{tabular}{ccc}
parameter & $r^2$ & $p$-value \\ \hline
$\log\beta$ & 0.08 & 0.98 \\
$\lambda_0$ & 0.12 & 0.84 \\
$\alpha$ & 0.12 & 0.8 \\
$\rho$ & 0.08 & 0.98 \\ \hline
\end{tabular}
\end{center}
\caption{Regression results after randomly permuting the vectors of 35 independent variables and tuple of 5 scoring dynamics parameters, $(\log\beta,\lambda_0,\alpha,\rho)$.}
\label{tab:randomize:regression}
\end{table}%

Finally, we perform a randomization test by randomly permuting the dependent variables across the associated features and repeating the original multivariate regression. This randomization destroys any natural correlation between the features and the dependent variable.
Table~\ref{tab:randomize:regression} shows the resulting coefficients of variation, none of which are statistically significant. These results further support the robustness of our original results.


\section{Player preference and competition balance}
\label{appendix:i}
When competitions are predictable they become less enjoyable. In professional sports this manifests itself as fans leaving a stadium well before the end of a game when one team is winning by such a large amount that there is little chance that the trailing team will make a comeback. 

In our model system, the same decision can occur for players themselves, who can effectively walk off the field by voluntarily exiting the competition early. For each of the competition types in our sample we calculated the competition dropout rate as
\begin{align}
\omega = \frac{1}{N}\sum_{i=1}^N\mathbbm{1}\{ \textrm{at least 1 player quits early} \},
\end{align}
where $N$ is the number of instances of the given type.

From the first 25 million games, we extracted a total of 4.1 million competitive type games that did not contain corrupt data. From these 4.1 million games we selected only those where at least one player left the game early. Using the remaining 1.9 million games we then tested for a correlation between the dropout rate $\omega$ and the overall balance $\beta$. If players prefer more balanced competitions, as $\beta$ increases (more ideal competitions), the dropout rate should decrease. A simple linear regression yields the equation $\ln \omega = 1.593 - 1.371\ln \beta$ ($r^2=0.43$, $p\ll0.001$, $t$-test). These results corroborate our hypothesis, illustrating that the more predictable the scoring dynamics of a competition (small $\beta$), the more likely at least one player will exit early. Quantitatively, this relationship predicts that increasing competition balance $\beta$ by a factor of 1.66 correlates with reducing the early exit probability $\omega$ by a factor of 2.

As a caveat, we note that there are several involuntary reasons a player may exit early, e.g., network issues, power loss, system error, being ``booted'' for excessive friendly fire, and several voluntary reasons unrelated to player engagement, e.g., to join friends in another game, to change competition types, etc. Most of these variables are inaccessible to us for analysis; however, we cannot conceive of a mechanistic relationship between most of these reasons and the scoring balance of a competition. Additional investigation may further illuminate the precise mechanism by which increase in $\beta$ produce decreased exit rates.


\bibliography{competition_refs}

\begin{table*}[tb]
\begin{center}
\begin{tabular}{c|llcl}
& feature & code & domain & description \\ \hline
\multirow{17}{*}{\begin{sideways}resources\end{sideways}} &loadout\_1	& R1 & $\{0,1\}$ & short range and medium range \\
&loadout\_2 & R2 & $\{0,1\}$ & low quality resources \\
&loadout\_3	 & R3 & $\{0,1\}$ & long range and grenades \\
&loadout\_4	 & R4 & $\{0,1\}$ & short and long range \\
&loadout\_5	 & R5 & $\{0,1\}$ & medium range \\
&vehicles\_revenant	& R6 & $\{0,1\}$ & lightly armored vehicle \\
&vehicles\_scorpion	& R7 & $\{0,1\}$ &  heavy tank vehicle \\
&vehicles\_mongoose	& R8 & $\{0,1\}$ &  unarmored vehicle \\
&vehicles\_ghost	& R9 & $\{0,1\}$ &  rapid attack vehicle \\
&weapons\_short	& R10 & $\{0,1\}$ & short range \\
&weapons\_medium	& R11 & $\{0,1\}$ & medium range \\
&weapons\_long	& R12 & $\{0,1\}$ & long range \\
&weapons\_grenades	& R13 & $\{0,1\}$ & grenade type \\
&weapons\_rocket	& R14 & $\{0,1\}$ & rocket launcher \\
&weapons\_unsc	& R15 & $\{0,1\}$ & high-quality only resources \\
&weapons\_covenent	& R16 & $\{0,1\}$ & low-quality only resources \\
&weapons\_both & R17 & $\{0,1\}$ & high- and low-quality resources \\
\hline
\multirow{2}{*}{\begin{sideways}skill\end{sideways}}  &TrueSkill matchmaking~ & S1	 & $\{0,1\}$ & equally skilled teams \\
&team size	& S2 & $\{0,1\}$ & 4- or 5-person teams \\
\hline
\multirow{12}{*}{\begin{sideways}environmental structure\end{sideways}} &map\_open & E1 & $\{0,1\}$ & open terrain \\
&map\_vertical & E2 & $\{0,1\}$ & vertical environment \\
&map\_circular & E3 & $\{0,1\}$ & circular terrain \\
&map\_varied & E4 & $\{0,1\}$ & no clear organizing principle \\
&map\_corridors & E5 & $\{0,1\}$ & indoor terrain \\
&map\_bases & E6 & $\{0,1\}$ & defensible positions \\
&map\_towers & E7 & $\{0,1\}$ & high ground \\
&map\_transporters & E8 & $\{0,1\}$ & teleporters, jump pads and vents \\
&map\_outdoor & E9 & $\{0,1\}$ & outdoor terrain \\
&map\_size\_small & E10 & $\{0,1\}$ & small or medium sized map \\
&map\_size\_large~~ & E11 & $\{0,1\}$ & large arena \\
&map\_size\_perim & E12 & $\mathbb{R}^{+}$ & perimeter of map, seconds required to run in game \\
\hline
\multirow{4}{*}{\begin{sideways}policies\end{sideways}} &rules\_noradar & P1 & $\{0,1\}$ & HUD radar is off \\
&rules\_noshields & P2 & $\{0,1\}$ & shield is off \\
&rules\_headshot & P3 & $\{0,1\}$ & headshot required for kill (SWAT rules) \\
&rules\_snipers & P4 & $\{0,1\}$ & sniper fight \\ \hline
\end{tabular}
\end{center}
\caption{Competition features, abbreviations and verbal descriptions, grouped in four categories: resources (R), skill (S), environmental structure (E), and policy (P).}
\label{tab:features}
\end{table*}%

\begin{table*}[tb!]
\begin{center}
\begin{tabular}{c|lccrr|c}
parameter & feature & $\theta$ & ~~std. error & ~~$t$ value & ~~$\Pr(>|t|)$ & $r^2$ \\ \hline
\multirow{12}{*}{$\log\beta$} & E5 & \hspace{0.8em}1.849 & 0.320  & 5.764 & $\ll0.001$ & \multirow{12}{*}{0.933} \\
& E1  & \hspace{0.8em}1.391 & 0.371 &  3.745 & $\ll0.001$ & \\
& E11 &  \hspace{0.8em}1.123 &  0.141 & 7.920 & $\ll0.001$ & \\
& S1  & \hspace{0.8em}0.822 & 0.034 & 23.828  & $\ll0.001$ & \\
& E3 & \hspace{0.8em}0.570 & 0.256  & 2.224 & 0.028 & \\  
& E9 &  \hspace{0.8em}0.481 &  0.076 & 6.265 & $\ll0.001$ & \\
& R10 & $-0.354$  & 0.134 &  $-2.642$ & 0.009 & \\  
& R8 & $-0.495$ & 0.215 & $-2.303$ &  0.023 & \\
& R15 & $-0.580$ & 0.233 & $-2.488$ & 0.014 & \\  
& E6  & $-0.813$ &  0.150  & $-5.414$ & $\ll0.001$ & \\
& E2 &  $-1.861$  & 0.252 &  $-7.375$ & $\ll0.001$ & \\
& E7  & $-2.126$  & 0.224  & $-9.467$ & $\ll0.001$ & \\ \hline
\multirow{17}{*}{$\lambda_{0}$} & E5 & \hspace{0.8em}0.082 & 0.008 &  9.966  & $\ll0.001$ & \multirow{17}{*}{0.955} \\
& E11 & \hspace{0.8em}0.059 & 0.003 & 16.344  & $\ll0.001$ & \\
& E1 & \hspace{0.8em}0.045 & 0.009 &  4.774 & $\ll0.001$ & \\
& E3 & \hspace{0.8em}0.029 & 0.006 &  4.437 & $\ll0.001$ & \\
& E9 & \hspace{0.8em}0.023 & 0.001 & 12.028  & $\ll0.001$ & \\  
& R10 &  \hspace{0.8em}0.008 & 0.003 &  2.478 & 0.014 & \\ 
& S1 &  \hspace{0.8em}0.005 & 0.001 &  6.010 & $\ll0.001$ & \\
& E4 & $-0.009$ & 0.004 & $-2.374$ & 0.019 & \\  
& R8 & $-0.011$&  0.005 & $-1.995$ & 0.048 & \\
& R13 & $-0.011$&  0.004 & $-2.266$ & 0.025 & \\
& E6 & $-0.011$ & 0.003 & $-2.845$ & 0.005 & \\ 
& R2 & $-0.015$ & 0.008 & $-1.873$ & 0.063 & \\
& R1 & $-0.021$ & 0.008 & $-2.680$ & 0.008 & \\ 
& R4 & $-0.030$ & 0.008 & $-3.797$ & $\ll0.001$ & \\   
& R15 & $-0.032$ & 0.006 & $-5.444$ & $\ll0.001$ & \\     
& E2 & $-0.081$ & 0.006 & $-12.448$  & $\ll0.001$ & \\
& E7  &  $-0.081$ & 0.005 & $-13.991$  & $\ll0.001$ & \\ \hline
\multirow{2}{*}{$\alpha$} & R12 & $-1.9\times10^{-5}$ & $8.1\times10^{-6}$ & $-2.449$  & 0.016 & \multirow{2}{*}{0.652} \\
& S1 & $-2.9\times10^{-6}$ & $1.7\times10^{-6}$ & $-1.692$ &  0.093 & \\ \hline
\multirow{14}{*}{$\rho$} & E7 & \hspace{0.8em}0.138 &  0.022 &  6.295 & $\ll0.001$ & \multirow{14}{*}{0.885} \\ 
& E2 & \hspace{0.8em}0.123  & 0.024 &  4.989 & $\ll0.001$ & \\
& R4 & \hspace{0.8em}0.070 & 0.030 &  2.299 & 0.023 & \\
& E6 & \hspace{0.8em}0.061 &  0.014 &  4.175 & $\ll0.001$ & \\
&R1 & \hspace{0.8em}0.053 &  0.030 &  1.734 & 0.085 & \\
& R15 & \hspace{0.8em}0.046 &  0.022 &  2.030 &  0.044 & \\  
& R8 & \hspace{0.8em}0.040 &  0.021  & 1.937 & 0.055 & \\
& E4 & \hspace{0.8em}0.031 &  0.015 &  2.018 & 0.046 & \\ 
& R3 & \hspace{0.8em}0.029 & 0.015 & 1.852 & 0.066 & \\
& R14 & $-0.030$ &  0.012 & $-2.366$ & 0.019 & \\
& E9 & $-0.036$ &  0.007 & $-4.775$ & $\ll0.001$ & \\ 
& S1 & $-0.055$ &  0.003 & $-16.413$  & $\ll0.001$ & \\
& E11 & $-0.089$ &  0.013 & $-6.410$ & $\ll0.001$ & \\
& E5 & $-0.095$ & 0.031 & $-3.020$ & 0.003 & \\ \hline
\end{tabular}
\end{center}
\caption{Ordered multivariate regression model coefficients for all standard (``slayer'') competitions regressed onto the estimated generative model parameters $\log\beta$, $\lambda_0$, $\alpha$, and predictability measure $\rho$.}
\label{tab:lin:reg}
\end{table*}%
\begin{table*}[htdp]
\begin{center}
\begin{tabular}{l|cccccc}
feature & df & \begin{sideways}Wilks\end{sideways} & \begin{sideways}approx.\ F\end{sideways} & \begin{sideways}num.\ df\end{sideways} & \begin{sideways}den.\ df\end{sideways} & $\Pr(>F)$ \\ \hline
R1 & 1 & 0.533 & \hspace{0.5em}21.617 & 4 & 99 & $\ll 0.001$ \\
R2 & 1 & 0.339 & \hspace{0.5em}48.147 & 4 & 99 & $\ll 0.001$ \\
R3 & 1 & 0.352 & \hspace{0.5em}45.541 & 4 & 99 & $\ll 0.001$ \\
R4 & 1 & 0.716 & \hspace{0.9em}9.802 & 4 & 99 & $\ll 0.001$ \\
R8 & 1 & 0.167 & 123.322 & 4 & 99 & $\ll 0.001$ \\
R10 & 1 & 0.302 & \hspace{0.5em}57.109 & 4 & 99 & $\ll 0.001$ \\
R11 & 1& 0.418 & \hspace{0.5em}34.459 & 4 & 99 & $\ll 0.001$ \\
R12 & 1 & 0.383 & \hspace{0.5em}39.799 & 4 & 99 & $\ll 0.001$ \\
R13 & 1 & 0.817 & \hspace{0.9em}5.536 & 4 & 99 & $\ll 0.001$ \\
S1 & 1 & 0.112 & 194.402 & 4 & 99 & $\ll 0.001$ \\
R15 & 1 & 0.224 & \hspace{0.5em}85.703 & 4 & 99 & $\ll 0.001$ \\
E1 & 1 & 0.455 & \hspace{0.5em}29.610 & 4 & 99 & $\ll 0.001$ \\
E2 & 1 & 0.358 & \hspace{0.5em}44.342 & 4 & 99 & $\ll 0.001$ \\
E3 & 1 & 0.606 & \hspace{0.5em}16.076 & 4 & 99 & $\ll 0.001$ \\
E4 & 1 & 0.811 & \hspace{0.9em}5.742 & 4 & 99 & $\ll 0.001$ \\
E5 & 1 & 0.246 & \hspace{0.5em}75.711 & 4 & 99 & $\ll 0.001$ \\
E6 & 1 & 0.399 & \hspace{0.5em}37.133 & 4 & 99 & $\ll 0.001$ \\
E7 & 1 & 0.842 & \hspace{0.9em}4.623 & 4 & 99 & \hspace{1.3em}0.001 \\
E9 & 1 & 0.401 & \hspace{0.5em}36.896 & 4 & 99 & $\ll 0.001$ \\
E11 & 1 & 0.239 & \hspace{0.5em}78.378 & 4 & 99 & $\ll 0.001$ \\ \hline
\end{tabular}
\end{center}
\caption{
MANOVA results of multiple multivariate regression model, providing a robustness check on the results given in Table~\ref{tab:lin:reg}.}
\label{tab:manova}
\end{table*}%

\begin{table*}[tb]
\begin{center}
\begin{tabular}{c|lcccr|c}
parameter & feature & $\theta$ & ~~std. error & ~~$t$ value & ~~$\Pr(>|t|)$ & $r^2$ \\ \hline
\multirow{18}{*}{$\log\beta$} & E5 & 1.803 & 0.229 & 7.867 & $\ll0.001$ & \multirow{18}{*}{0.933} \\
& E1 & 1.320 & 0.228 & 5.779 & $\ll0.001$ & \\
& E11 & 1.126 & 0.124 & 9.029 & $\ll0.001$ & \\
& S1 & 0.822 & 0.034 & 24.153  & $\ll0.001$ & \\
& E3 & 0.480 & 0.122 & 3.919 & $\ll0.001$ & \\
& E9 & 0.479 & 0.069 & 6.888 & $\ll0.001$ & \\
& R13 & 0.154 & 0.069 & 2.243 & 0.027 & \\  
& R14 & 0.119 & 0.074 & 1.598 & 0.113 & \\   
& R1 & -0.322 & 0.054 & -5.952 & $\ll0.001$ & \\
& R3 & -0.232 & 0.092 & -2.505 & 0.013 & \\ 
& R12 & -0.310 & 0.110 & -2.822 & 0.005 & \\
& R10 & -0.367 & 0.113 & -3.232 & 0.001 & \\ 
& R8 & -0.472 & 0.181 & -2.596 & 0.01 & \\ 
& R4 & -0.504 & 0.062 & -8.081 & $\ll0.001$ & \\
& R15 & -0.644 & 0.092 & -6.931 & $\ll0.001$ & \\
& E6 & -0.827 & 0.130 & -6.353 & $\ll0.001$ & \\
& E2 & -1.860 & 0.207 & -8.957 & $\ll0.001$ & \\
& E7 & -2.093 & 0.193 & -10.840  & $\ll0.001$ & \\ \hline
\multirow{18}{*}{$\lambda_0$} & E5 & 0.084 & 0.006 & 13.770  & $\ll0.001$ & \multirow{18}{*}{0.954} \\
& E11 & 0.061 & 0.002 & 20.759 & $\ll0.001$ & \\
& E3 & 0.029 & 0.003 &  8.648 & $\ll0.001$ & \\
& E9 & 0.024 & 0.001 & 12.383  & $\ll0.001$ & \\
& R10 & 0.008 & 0.003 &  2.794 & 0.006 & \\
& R3 & 0.005 & 0.002 &  2.080 & 0.039 & \\
& S1 & 0.005 & 0.001 &  6.085 & $\ll0.001$ & \\
& E1 & 0.048 & 0.005 &  8.880 & $\ll0.001$ & \\
& R13 & -0.009 & 0.002 & -3.979 & $\ll0.001$ & \\
& E4 & -0.008 & 0.002 & -3.178 & 0.001 & \\ 
& R8 & -0.011 & 0.004 & -2.467 & 0.015 & \\ 
& E6 & -0.012 & 0.003 & -3.860 & $\ll0.001$ & \\
& R2 & -0.015 & 0.005 & -2.939 &0.004 & \\ 
& R1 & -0.022 & 0.005 & -4.191 & $\ll0.001$ & \\
& R4 & -0.031 &  0.005 & -5.852 & $\ll0.001$ & \\
& R15 & -0.034 & 0.004 & -8.469 & $\ll0.001$ & \\
& E7 & -0.080 & 0.004 & -16.695  & $\ll0.001$ & \\
& E2 & -0.081 & 0.005 & -14.457 & $\ll0.001$ & \\ \hline
\end{tabular}
\end{center}
\caption{Ordered multivariate regression model coefficients for all standard (``slayer'') competitions regressed onto $\log \beta$, $\lambda_0$, selected via stepwise AIC, providing a second check on the robustness of the results in Table~\ref{tab:lin:reg}.}
\label{tab:aic:a}
\end{table*}

\begin{table*}[tb]
\begin{center}
\begin{tabular}{c|lcccr|c}
parameter & feature & $\theta$ & ~~std. error & ~~$t$ value & ~~$\Pr(>|t|)$ & $r^2$ \\ \hline
\multirow{15}{*}{$\rho$} & E7 &  \hspace{0.8em}0.124 &  0.010  & \hspace{0.8em}11.934  & $\ll0.001$ & \multirow{15}{*}{0.882} \\
& E2 & \hspace{0.8em}0.111 &  0.011  & \hspace{1.3em}9.943  & $\ll0.001$ & \\
& R4 & \hspace{0.8em}0.067 &  0.010 &  \hspace{1.3em}6.444 & $\ll0.001$ & \\
& E6  & \hspace{0.8em}0.052 &  0.005  & \hspace{1.3em}8.998 & $\ll0.001$ & \\
& R1 &  \hspace{0.8em}0.049  & 0.010  & \hspace{1.3em}4.958 & $\ll0.001$ & \\
& R8 & \hspace{0.8em}0.046 &  0.016  & \hspace{1.3em}2.779 & 0.006 & \\ 
& R15  & \hspace{0.8em}0.045 &  0.006 &  \hspace{1.3em}7.335 & $\ll0.001$ & \\
& E4 &  \hspace{0.8em}0.039 &  0.007  & \hspace{1.3em}5.456 & $\ll0.001$ & \\
& R2 & \hspace{0.8em}0.037 &  0.010  & \hspace{1.3em}3.533 & $\ll0.001$ & \\
& R3 & \hspace{0.8em}0.027 &  0.008  & \hspace{1.3em}3.420 & $\ll0.001$ & \\
& E9  &  $-0.034$  & 0.006 & \hspace{0.5em}$-4.912$ & $\ll0.001$ & \\
& R14  &  $-0.036$  & 0.006 & \hspace{0.5em}$-5.971$ & $\ll0.001$ & \\
& S1 & $-0.055$ &  0.003 & $-16.763$  & $\ll0.001$ & \\
& E5  &  $-0.076$  & 0.010 & \hspace{0.5em}$-7.429$ & $\ll0.001$ & \\
& E11 &  $-0.081$ &  0.006 & $-12.389$  & $\ll0.001$ & \\ \hline
\end{tabular}
\end{center}
\caption{Ordered multivariate regression model coefficients for all standard (``slayer'') competitions regressed onto $\rho$ selected via stepwise AIC, providing a second check on the robustness of the results in Table~\ref{tab:lin:reg}.}
\label{tab:aic:b}
\end{table*}

\begin{table*}[tb]
\begin{center}
\begin{tabular}{c|lcccr|c}
parameter & feature & ~~$\theta$ ($\times10^{-5}$) & ~~std. error ($\times10^{-6}$) & ~~$t$ value & ~~$\Pr(>|t|)$ & $r^2$ \\ \hline
\multirow{10}{*}{$\alpha$} & R3 & \hspace{0.8em}1.570 & 2.583 &  \hspace{0.8em}6.077 & $\ll0.001$ & \multirow{10}{*}{0.637} \\
& R11 & \hspace{0.8em}1.446 & 3.328 &  \hspace{0.8em}4.345 & $\ll0.001$ & \\ 
& R2 & \hspace{0.8em}1.432 & 2.965 &  \hspace{0.8em}4.832 & $\ll0.001$ & \\
& E5 & \hspace{0.8em}1.105 & 2.114 &  \hspace{0.8em}5.226 & $\ll0.001$ & \\
& E3 & \hspace{0.8em}0.454 & 2.368  & \hspace{0.8em}1.918 &  0.057 & \\  
& S1&  $-0.294$ & 1.689 & $-1.746$ &  0.083 & \\
& R1 & $-0.470$ & 2.529 & $-1.859$ &  0.065 & \\
& R15  & $-1.591$ & 2.583 & $-6.157$ & $\ll0.001$ & \\
& R8 & $-1.868$ & 7.159 & $-2.609$ &  0.010 & \\
& R12 & $-2.551$ & 2.538 & $-10.053$  & $\ll0.001$ & \\ \hline
\end{tabular}
\end{center}
\caption{Ordered multivariate regression model coefficients for all standard (``slayer'') competitions regressed onto $\alpha$ selected via stepwise AIC, providing a second check on the robustness of the results in Table~\ref{tab:lin:reg}.}
\label{tab:aic:c}
\end{table*}